\documentclass[12pt]{article}

\usepackage{natbib}
\usepackage[nottoc,notlof,notlot]{tocbibind} 

\bibpunct{(}{)}{;}{a}{}{;}
\usepackage{breakcites}
\usepackage[breaklinks=true]{hyperref}
\hypersetup{
    colorlinks=true,
    linkcolor=blue,
    citecolor=blue,
    urlcolor=blue
}
\usepackage{graphicx}
\graphicspath{ {images/} }

\usepackage{xcolor}
\usepackage{algorithm,algorithmic}

\usepackage{amsmath}
\usepackage{bbm}
\usepackage{amsfonts}%
\usepackage{amssymb}%
\usepackage{amsthm}
\usepackage{geometry} % to change the page dimensions
\usepackage{graphicx} % support the \includegraphics command and options
\usepackage{framed}
\usepackage{caption}
\usepackage{subcaption}

\usepackage{booktabs} % for much better looking tables
\usepackage{array} % for better arrays (eg matrices) in maths
\usepackage{paralist} % very flexible & customisable lists (eg. enumerate/itemize, etc.)
\usepackage{verbatim} % adds environment for commenting out blocks of text & for better verbatim
\usepackage{lipsum}
\usepackage{setspace}

\usepackage{lipsum}
\usepackage[singlelinecheck=false]{caption}
\usepackage{tabularx}

\usepackage{multirow}

\title{\setstretch{1} Panel Data with Unknown Clusters}

\author{Yong Cai\thanks{Department of Economics, Northwestern University. E-mail: \url{yongcai2023@u.northwestern.edu}. I am grateful to Ivan Canay for extensive guidance on this project. I would also like to thank Eric Auerbach, Grant Goehring, Joel Horowitz and Isaac Loh for helpful comments and suggestions.}% This research was supported in part through the computational resources and staff contributions provided for the Quest high performance computing facility at Northwestern University which is jointly supported by the Office of the Provost, the Office for Research, and Northwestern University Information Technology.}
}
 \date{\parbox{\linewidth}{\centering%
  \today\endgraf}} % Activate to display a given date or no date (if empty),
\theoremstyle{definition}
\newtheorem{definition}{Definition}
\newtheorem{theorem}{Theorem}
\newtheorem{lemma}{Lemma}

\newtheorem{corollary}{Corollary}
\newtheorem{assumption}{Assumption}

\newtheorem{remark}{Remark}

\begin{document}
\maketitle
%\centerline{\large PRELIMINARY AND INCOMPLETE}
\begin{abstract} \setstretch{1}\noindent
Clustered standard errors and approximate randomization tests are popular inference methods that allow for dependence within observations. However, they require researchers to know the cluster structure ex ante. We propose a procedure to help researchers discover clusters in panel data. Our method is based on thresholding an estimated long-run variance-covariance matrix and requires the panel to be large in the time dimension, but imposes no lower bound on the number of units. We show that our procedure recovers the true clusters with high probability with no assumptions on the cluster structure. The estimated clusters are independently of interest, but they can also be used in the approximate randomization tests or with conventional cluster-robust covariance estimators. The resulting procedures control size and have good power. 
\end{abstract}

\section{Introduction}

Consider the following regression with panel data:
\begin{equation*}
	Y_{it} = X_{it}'\beta + U_{it} \quad , \quad E[X_{it}U_{it}] = 0
\end{equation*}
where a researcher wants to conduct inference on $\beta$. If the researcher is concerned about correlations between $U_{it}$ and $U_{i't'}$, it is frequently helpful to group units into independent clusters. These independent clusters can then be used to construct cluster-robust covariance estimators (CCE) as in \cite{lz1986}, or for approximate randomization tests as in \cite{crs2017} and \cite{ccks2021}. 

However, cluster assignments are rarely known ex ante. In many contexts, multiple levels of clustering are plausible. For example, with the American Community Survey, researchers have the choice of clustering at the individual, county or state level and the appropriate level of clustering is not always obvious. In other situations, researchers may not be able to identify their desired level of clustering. For example, in the presence of peer effects, researchers might like to cluster their observations along friend groups since this is the level at which spillover occurs. However, unless researchers observe friendship networks, it would not be possible to cluster at this chosen level. 

Clustering at the correct level is important for inference. It is well known that ignoring cluster dependence -– in other words, clustering at too fine a level -- leads to tests with excessive type I errors. \cite{bdm2004} and \cite{cgm2008}, for example, find nominal size find that neglecting dependence can lead to type I that exceeds their nominal size by as much as 10 times. Conversely, excessively coarse levels of clustering bring their own problems. Firstly, coarse clusters tend to be few in numbers. A large body of work show that confidence intervals based on the cluster-robust standard errors tend to under-cover when the number of clusters is small (see \cite{mhe2008} and \cite{cgm2008} among others), leading to poor size control. Secondly, setting aside under-coverage issues, excessively coarse levels of clustering cause tests to have poor power (\cite{aaiw17}), since the researcher assumes less information than they actually have. Despite the importance of these issues, there has been limited theoretical guidance on choosing the appropriate level of clustering.

We propose a procedure to help researchers discover clusters in the panel data. Our method is based on thresholding an estimated long-run variance-covariance matrix and requires the panel to be large in the time dimension, but imposes no lower bound on the number of units. We show that the procedure recovers the true clusters with high probability with few assumptions on the cluster structure. We believe the estimated clusters can be of independent interest to researchers. However, they can also directly be used in the approximate randomization tests of \cite{crs2017} or in tests based on cluster-robust covariance estimators. We show that doing so leads to tests which control size in asymptotic frameworks that takes the number of clusters to be fixed or growing to infinity. 

Our paper is similar in spirit to tests for level of clustering, which aims to provide a robustness checks for researchers who have already chosen a given level of clustering. However, these methods are designed to test between two specified nested levels of clustering and are not suitable for discovering the cluster structure. Our paper also relates to the extensive literature on panel data with interactive fixed effects, which effectively assumes that cluster dependence takes factor structure. We are unable to accommodate the type of endogeneity allowed in this literature, but we allow for richer patterns of dependence while maintaining ease of computation. 

Our paper is most similar to \cite{bcl2020}, which provides a method for inference in panel data when clusters are unknown, but when correlation across units are sparse. Both our methods are based on thresholding a long-run variance-covariance matrix. The method of \cite{bcl2020} even when in the absence of cluster structures, since they rely on sparsity in the dependence structure between units. In the cluster setting, their sparsity assumption translates into many-cluster asymptotics, which may not be realistic in all applications. 
Unlike their method, ours is able to accommodate small-cluster asymptotics. Furthermore, we provide a novel cluster-recovery result. Simulations also suggest that our method leads to tests which are more powerful. Finally, \cite{aaiw17} advocates a design-based approach on the issue of clustering, arguing that dependence between units is neither necessary nor sufficient for clustering standard errors. We argue that our method is useful even for researchers who takes such a perspective. We expand on the aforementioned points in section \ref{section--existinglit}.

The rest of this paper is organized as follows. Section \ref{section--model} presents our model and assumptions. The proposed method as well as our theoretical results are contained in section \ref{section--method}. Section \ref{section--existinglit} relates our paper to existing literature. Section \ref{section--simulations} presents results from Monte Carlo simulation. Section \ref{section--conclusion} concludes. Proofs are contained in the appendices. 

\section{Model and Assumptions}\label{section--model}

In this section we discuss the assumptions which are needed for cluster recovery and inference. The most onerous assumption of the method is that it requires panel data that is large in the time dimension. Beyond that, we require the covariates and error terms to have tails that are sufficiently thin, and also for dependency across time to decay quickly enough. Our assumptions are generally standard in the panel data literature. 

We work with the usual linear model:
\begin{assumption}[Model] \label{assumption--model}
	Consider the model
	\begin{equation*}
	Y_{i,t} = X_{i,t}'\beta + U_{i,t} \quad , \quad E[X_{i,t}U_{i,t}] = 0
	\end{equation*}
	Suppose also that for all $N, T \in \mathbb{N}$, there exists $\underline{\lambda} \in \mathbb{R}$ so that
	\begin{equation*}
		\lambda_\text{min} \left( \frac{1}{NT}\sum_{i=1}^N\sum_{t=1}^T E\left[X_{i,t}X_{i,t}'\right]\right) \geq \underline{\lambda}
	\end{equation*}
	where $\lambda_\text{min}(M)$ is the smallest singular value of the matrix $M$.  
%	so that for all $N$ and $T$:
%	\begin{equation*}
%		\beta = \left(\frac{1}{NT}\sum_{i=1}^N\sum_{t=1}^T E[X_{i,t}X_{i,t}']\right)^{-1} \frac{1}{NT}\sum_{i=1}^N\sum_{t=1}^T E[X_{i,t}Y_{i,t}']
%	\end{equation*}
\end{assumption}
The lower bound on the singular values of the expected Gram matrix is a common alternative for the full rank assumption when working with independent but not identical clusters. 

Our method learns cluster structure from what we will call the long-run correlation matrix. For this matrix to be estimable, we assume strong mixing and stationarity at the unit level. %These conditions are not unlike those found in \cite{bcl2020}. It is also common in the literature on inference with few clusters (see for instance \cite{crs2017} and \cite{bch2011}). 
\begin{definition}
	Define the $\alpha$-mixing coefficient (for stationary random variables) as
	\begin{equation*}
		\alpha(h) = \sup_{A \in \mathcal{F}^0_{-\infty} , B \in \mathcal{F}_h^\infty } \lvert P(A)P(B) - P(A \cup B) \rvert
	\end{equation*}
	where $\left(\mathcal{F}^0_{-\infty} , \mathcal{F}_h^\infty \right)$ are the $\sigma$-algebras generated by $\{\mathbb{X}_t, \mathbb{U}_t\}^0_{t=-\infty}$ and  $\{\mathbb{X}_t, \mathbb{U}_t\}^\infty_{t=h}$ respectively. 
\end{definition}

Our strong mixing assumption is that:
\begin{assumption}[Strong Mixing]\label{assumption--strongmixing}
	Suppose $\{\mathbb{X}_t, \mathbb{U}_t\}$ are stationary, and there exists $C_1 > 0$ and $\kappa > 0$ such that $\alpha(h) \leq \exp(-C_1h^\kappa)$. 
\end{assumption}

In order to achieve a sufficiently fast rate of convergence, we also need to control higher moments of the scores. In particular, we assume that they decrease fast enough to satisfy Bernstein's condition:
\begin{assumption}[Bernstein's Condition]\label{assumption--bernstein}
	Suppose there exists $C_2 > 0$ and $M$ such that for all $h \in \mathbb{Z}_+$ and $i, j \in \mathbb{N}$, we have that for all $k \geq 1$,  
	\begin{equation*}
		E\left[\lvert W_{i,t}W_{j,t-h} \rvert^k \right] \leq C_2^{k-2}k!E[W_{i,t}^2W_{j,t-h}^2]~,
	\end{equation*}
	where $W_{it} \in \left\{X^{(1)}_{i,t}U_{i,t}, ..., X^{(p)}_{i,t}U_{i,t} \right\}$ and $W_{j,t-h} \in \left\{X^{(1)}_{j,t-h}U_{j,t-h}, ..., X^{(p)}_{j,t-h}U_{j,t-h} \right\}$.
\end{assumption}
Assumptions \ref{assumption--strongmixing} and \ref{assumption--bernstein} -- or their analogues -- are frequently seen in the panel data literature. See for instance \cite{bcl2020} or \cite{bm2015}. 

Our next assumption restricts heterogeneity across individuals:
\begin{assumption}[Uniformity Conditions]\label{assumption--uniformity}
Suppose there exists $M_2$ and $M_k$ for some $k \geq 4$ such that for all $h \in \mathbb{Z}_+$ and $i, j \in \mathbb{N}$
	\begin{align*}
		E[W_{i,t}^2W_{j,t-h}^2] \leq M_2 \quad \mbox{and} \quad E[W_{i,t}^kW_{j,t-h}^k] \leq M_k 
	\end{align*}
	for all $i$, $j$, $h$ , $W_{it} \in \left\{X^{(1)}_{i,t}U_{i,t}, ..., X^{(p)}_{i,t}U_{i,t} \right\}$ and $W_{j,t-h} \in \left\{X^{(1)}_{j,t-h}U_{j,t-h}, ..., X^{(p)}_{j,t-h}U_{j,t-h} \right\}$.
\end{assumption}
In particular, we require that the variance and covariance across individuals to be bounded. As such, while individuals can be different, we cannot have a few individuals dominating the regression estimates. 

The next assumption concerns the existence of a cluster structure. 
\begin{definition}
	Define:
	\begin{align*}
		\sigma_{i,j}^{a,b} = E[X_{i,t}^{(a)}U_{i,t}X_{j,t}^{(b)}U_{j,t}] + \sum_{h=1}^{\infty} 	E[X_{i,t}^{(a)}{U}_{i,t}X_{j,t-h}^{(b)}{U}_{j,t-h}] + E[X_{i,t-h}^{(a)}{U}_{i,t-h}X_{j,t}^{(b)}{U}_{j,t}] ~.
	\end{align*}		
	where $X_{i,t}^{(a)}$ refers to the $a^{\text{th}}$ component of the vector $X_{i,t}$. 
	Let the long-run covariance between two individuals be denoted
	\begin{align*}
			\sigma_{i,j} = \sum_{a = 1}^p \sum_{b = 1}^p \left\lvert \sigma_{i,j}^{a,b} \right\rvert ~.
	\end{align*}	
	Define $\Sigma$ to be the $N\times N$ matrix with $\sigma_{i,j}/\sqrt{\sigma_{i,i} \sigma_{j,j}}$ in its $(i,j)^\text{th}$ entry. 
%		\begin{align*}
%				\left[\Sigma(t) \right]_{i,j} = \frac{1}{T}\sum_{t=1}^{T} E[U_{i,t}U_{j,t}] + \sum_{h = 1}^T \frac{1}{T-h} \sum_{t=h+1}^{T} E[{U}_{i,t}{U}_{j,t-h}] + E[{U}_{i,t-h}{U}_{j,t}]
%		\end{align*}
\end{definition}

\begin{assumption}[Cluster Structure]\label{assumption--clusterstructure}
	Suppose that each individual $i$ belongs to one of $q$ clusters, where $q$ is unknown. Let $g(i): [N] \to [q]$ denote the function that maps $i$ to its cluster. Further, suppose that 
	\begin{enumerate}
		\item[a.] $\sigma_{ij} = 0$ if $i$ and $j$ belong to different cluster
		\item[b.] For each $i$, $|g(i)| > 1$ implies that there exists at least one $j \neq i$ s.t. $g(j) = g(i)$ and $\sigma_{ij} \neq 0$.
		\item[c.] There exists $0 < \underline{\sigma} \leq \overline{\sigma} < \infty$ such that for all $i,j \in \mathbb{N}$, $\sigma_{i,j} \neq 0 \Rightarrow \underline{\sigma} \leq\sigma_{ij} \leq \overline{\sigma}$. 
	\end{enumerate}
	%\begin{align*}
	%	\sigma_{ij} = \begin{cases}
	%		\sigma_k & \mbox{ if $i$ and $j$ both belong to cluster $k$}, \\
	%		0 &\mbox{ otherwise.}
	%	\end{cases}
	%\end{align*}
\end{assumption}

In other words, when the individuals are sorted according to their clusters, we have that:
\begin{equation*}
\Sigma =  \begin{pmatrix}
\Sigma_1 & 0 & 0 &\cdots & 0 \\
0 & \Sigma_2 & 0 & \cdots & 0 \\
\vdots & \vdots & \vdots & & \vdots \\
0 & 0 & 0 & \cdots & \Sigma_q
\end{pmatrix} %\quad \mbox{ and } \quad 	
%\Sigma_k =  \begin{pmatrix}
%		0	& \sigma_k & \sigma_k  & \cdots & \sigma_k \\
%		\sigma_k & 0 & \sigma_k  & \cdots & \sigma_k  \\
%		\vdots & \vdots & \vdots & & \vdots  \\
%		\sigma_k & \sigma_k & \sigma_k  & \cdots & \sigma_k \\
%		\sigma_k & \sigma_k & \sigma_k  & \cdots & 0 \\
%		\end{pmatrix}
\end{equation*}
where each $\Sigma_g$ is either a scalar, or has a non-zero entry in every row. 

The above assumption does not restrict cluster structure since we make no assumption on $q$. In particular, any pattern of correlation between units is permitted if we set $q = 1$. As will become clear in section \ref{section--mainresults}, we do not need any assumption on $q$ to recover the cluster structure. However, we will introduce assumptions on $q$ for inference. 

Where the assumption has bite is in restricting $\sigma_{i,j}$. The lower bound is similar to the beta-min condition seen in the LASSO literature. It rules out covariance terms that are arbitrarily close to $0$, which would be difficult to estimate. 

In addition, we require the $\sigma_{i,j}$ to be bounded above. This ensures that the estimate of the long-run correlation matrix $\Sigma$ is well-behaved. However, we can also threshold a covariance matrix rather than a correlation matrix, in which case the upper bound would be unnecessary. 

%In comparison, \cite{bcl2020} require a sparsity assumption for their method. They do not require the long-run variance-covariance matrix to have cluster structure. However, in a cluster context, their assumption implies many small clusters. It is incompatible with having a few large clusters, a situation that is of practical importance in empirical work.  

Lastly, we note that our goal is only to recover the cluster structure up to relabeling. To be precise, we define the following:
\begin{definition}
	We say that cluster structure $g$ is equivalent to $\tilde{g}$ if there exists permutation $\pi$ such that $g([N]) = \pi \tilde{g}([N])$. We write $g \cong \tilde{g}$. 
\end{definition}

In other words, we seek $\hat{g}$ so that $\hat{g} \cong g$ with high probability. 

\section{The Proposed Method}\label{section--method}

In this section we discuss our proposed method for recovering the clusters and for performing inference. The method involves $2$ tuning parameters. We provide heuristics for choosing these parameters in section \ref{section--crossvalidation}. 

\subsection{Cluster Recovery}

We first define our estimator for the long-run correlation matrix. The estimator is based on the heteroskedasticity and autocorrelation consistent (HAC) estimator of \cite{nw1994} and involves a bandwidth (tuning) parameter.

\begin{definition}[Bartlett Kernel]\label{definition--Bartlett}
	We define the Bartlett Kernel to be $\omega: \mathbb{Z}_+ \times \mathbb{Z}_+ \to \mathbb{R}$,
	\begin{align*}
		\omega(h, L) = \begin{cases}
			\frac{L - h}{L} & \mbox{ if } h \leq L \\
			0 & \mbox{ otherwise.}
		\end{cases}
	\end{align*}
\end{definition}

\begin{definition}
	Given the bandwidth parameter $L \in \mathbb{Z}_+$, define:
	\begin{align}\label{eqn--sigmahat-def}
		\begin{aligned}
		\hat{\sigma}^{a,b}_{i,j} & = \frac{1}{T}\sum_{t=1}^{T} X_{i,t}^{(a)}\hat{U}_{i,t}X_{j,t}^{(b)}\hat{U}_{j,t} \\
		& + \sum_{h = 1}^L \frac{\omega(h, L)}{T-h} \sum_{t=h+1}^{T} X_{i,t}^{(a)}\hat{U}_{i,t}X_{j,t-h}^{(b)}\hat{U}_{j,t-h} \\
		& + \sum_{h = 1}^L \frac{\omega(h, L)}{T-h} X_{i,t-h}^{(a)}\hat{U}_{i,t-h} X_{j,t}^{(b)}\hat{U}_{j,t-h}
		\end{aligned}
	\end{align}
	and let 
	\begin{equation*}
		\hat{\sigma}_{i,j} = \sum_{a = 1}^{p} \sum_{b = 1}^p \left\lvert \hat{\sigma}_{i,j}^{a,b} \right\rvert
	\end{equation*}
	Let the estimator of $\Sigma$, denoted $\hat{\Sigma}$, be the $N\times N$ matrix with $\hat{\sigma}_{i,j}/\sqrt{ \hat{\sigma}_{i,i} \hat{\sigma}_{j,j} }$ in its $(i,j)^\text{th}$ entry.
\end{definition}

We propose to estimate the cluster structure using algorithm \ref{algorithm--clusterrecovery}. In words, the algorithm estimates $\Sigma$ entry-by-entry using a HAC-based estimator. Given $\hat{\Sigma}$, we set correlations that are smaller than $T^{\eta - 1/2}$ to $0$, because these links are likely spurious. Using the remaining links, we can then group individuals who are correlated into the same cluster. 

\begin{algorithm}[h!]
\caption{Adaptive Clustering}\label{algorithm--clusterrecovery}\small
\begin{algorithmic}[1]
	\STATE Fix the tuning parameters $L \in \mathbb{Z}_+$ and $0 < \eta < \frac{1}{2}$. 
	\STATE Compute the full sample regression of $Y_{it}$ on $X_{it}$ to obtain $\hat{\Sigma}$.
	\STATE Remove links in $\hat{G}$ for which $|\hat{\sigma}_{i,j}| < T^{\eta-1/2} $. 
	\STATE Assign all $i,j$ such that $|\hat{\sigma}_{i,j}| \neq 0$ to the same cluster. 
	\STATE Label the resulting clusters from $1$ to $\hat{q}$.
\end{algorithmic}
\end{algorithm}

\begin{remark}
	We state our result for the Bartlett kernel. Our obtained rates of convergence rely on the fact that partial sums of the Bartlett kernel is bounded by $\log_2(T)$. Other kernels can be used but they might lead to different rates of convergence. 
\end{remark}

\subsection{Inference}

Using the estimated clusters, there are two possible methods for performing inference, depending on the assumption that is made on $q$. When $q$ is small, the approximate randomization tests of \cite{crs2017} controls size. It's main drawback is that $\hat{\beta}$ needs to be estimable cluster-by-cluster. On the other hand, if $q$ is large, inference using the conventional clustered covariance estimator errors will have good properties, though this method is not compatible with small $q$. 

\subsubsection{Approximate Randomization Tests}

Approximate Randomization Tests (ARTs) are appropriate when there are few clusters. In this subsection we explain how to test linear hypotheses using ARTs. For ease of exposition, we discuss the case with a single linear restriction, though the test also accommodates tests of multiple linear restrictions. The hypothesis of interest is:
\begin{equation}\label{equation--hypothesis}
	H_0: r'\beta = \lambda \quad \mbox{against} \quad H_1: r'\beta \neq \lambda
\end{equation}
The test is based on cluster-by-cluster estimates of $\beta$, which we denote $\{\hat{\beta}_j\}_{j=1}^{\hat{q}}$. Define:
  \begin{equation*}
      S_T = \left(r'\hat{\beta}_1 - \lambda, \, ...\, , r'\hat{\beta}_{\hat{q}} - \lambda \right)'
  \end{equation*} 
Our test statistic is:
\begin{equation*}
 	R(S_T) = \frac{\sqrt{\hat{q}} \left\lvert \sum_{j=1}^{\hat{q}} r'\hat{\beta}_j - \lambda \right\rvert }{\sum_{j=1}^{\hat{q}} \left(r'\hat{\beta}_j - r'\bar{\hat{\beta}} \right)^2} \quad , \quad \bar{\hat{\beta}} = \frac{1}{\hat{q}} \sum_{j=1}^{\hat{q}} \hat{\beta}_j~.
\end{equation*}
This is the familiar $t$-statistic that takes $S_T$ as the ``raw data". Intuitively, $R(S_T)$ is large if the $r\hat{\beta}_j$'s are far from $\lambda$ and small otherwise. 

Next, denote by $\mathbf{G}$ the group of $\hat{q} \times 1$ sign changes. $\mathbf{G}$ can be identified with the set of $g \in \{-1, 1\}^{\hat{q}}$ so that:
\begin{equation*}
	gS_T = \begin{pmatrix}
		g_1 \cdot (r'\hat{\beta}_1 - \lambda) \\
		\vdots \\
		g_{\hat{q}} \cdot (r'\hat{\beta}_{\hat{q}} - \lambda)		
	\end{pmatrix}
\end{equation*}
As we elaborate in section \ref{section--mainresults}, the basis for our randomization test is that asymptotically, $gS_T$ has the same distribution as $S_T$. Hence, the set of $R(gS_T)$ provides a valid reference distribution for the $R(S_T)$. The test therefore rejects the null hypothesis when $R(S_T)$ takes on extreme values relative to $R(gS_T)$. It proceeds as follows. 

Define $M = |\mathbf{G}| = 2^{\hat{q}}$ and let:
\begin{equation*}
R^{(1)}(S_T) \leq R^{(2)}(S_T) \leq ... \leq R^{M}(S_T)
\end{equation*}
be the ordered values of $R(gS_T)$ as $g$ varies in $\mathbf{G}$. For a fixed nominal level $\alpha$, let $k$ be defined as
\begin{equation*}
k = \lceil (1-\alpha) M \rceil~,
\end{equation*}
where $\lceil x \rceil$ denotes the smallest integer greater or equal to $x$. In addition, define:
\begin{equation}\label{eqn--defineM}
\begin{aligned}
M^+(S_T) & = \sum_{j=1}^{M} I\{R^{(j)}(S_T) > R^{(k)}(S_T)\} \\
M^0(S_T) & = \sum_{j=1}^{M} I\{R^{(j)}(S_T) = R^{(k)}(S_T)  \}
\end{aligned}
\end{equation}
and set
\begin{equation}\label{eqn--definea}
a(S_T) = \frac{M\alpha - M^+(S_T)}{M^0(S_T)}~.
\end{equation}
We can then define the randomization test as:
\begin{align} \label{eqn--randtest}
\phi^{ART}_T = \begin{cases}
1 & \mbox{ if }R(S_T) > R^{(k)}(S_T), \\
a(S_T) & \mbox{ if }R(S_T) = R^{(k)}(S_T), \\
0 & \mbox{ otherwise.}
\end{cases}
\end{align}
In words, this test rejects the null hypothesis with certainty when $R(S_T) > R^{(k)}(S_T)$. When $R(S_T) = R^{(k)}(S_T)$, it rejects the null hypothesis with probability $a(S_T)$. The test does not reject when $R(S_T) < R^{(k)}(S_T)$.

\begin{remark}
As with standard randomization tests, $|\mathbf{G}|$ may sometimes be too large so that computation of $\{R(gS_T)\}_{g \in \mathbf{G}}$ becomes onerous. In these instances, it is possible to replace $\{R(gS_T)\}_{g \in \mathbf{G}}$ with a stochastic approximation. Formally, for a given $B \in \mathbb{Z}_+$, let
\begin{equation*}
	\hat{\mathbf{G}} = \left\{ g^1, ... ,g^B \right\}
\end{equation*}
where $g^1$ is the identity transformation and $g^2, ..., g^B$ are independent draws from Uniform($\mathbf{G}$). Using $\hat{\mathbf{G}}$ instead of ${\mathbf{G}}$ in equation (\ref{eqn--randtest}) does not affect validity of our results in section \ref{section--mainresults}.
\end{remark}

\begin{remark}
	Our test is possibly randomized. A deterministic but conservative version of the test can be implemented by rejecting the null hypothesis if and only if $R(S_T) > R^{(k)}(S_T)$. In the spirit of restricting researcher degree-of-freedom, this is also the version of the test that we implement in the Monte Carlo simulations of section \ref{section--simulations}. With appropriately chosen tuning parameters, the test has good power. 
\end{remark}

In sum, we propose to conduct inference by treating the estimated clusters as the true clusters and applying the usual ART. We summarize the implementation procedure in the algorithm below:
\begin{algorithm}[h!]\small
\caption{Approximate Randomization Test}\label{algorithm--art}
\begin{algorithmic}[1]
  \STATE Given the estimated clusters $\hat{g}$ from algorithm \ref{algorithm--clusterrecovery}, compute the cluster-by-cluster estimates of $\beta$, denoted $\{\hat{\beta}_1, ..., \hat{\beta}_{\hat{q}}\}$.
  \STATE Define:
    \begin{equation*}
        S_T = \left(\sqrt{n_1}(r'\hat{\beta}_1 - \lambda), \, ...\, , \sqrt{n_{\hat{q}}}(r'\hat{\beta}_{\hat{q}} - \lambda) \right)'
    \end{equation*} 
    and compute the test statistic:
	\begin{equation*}
	 	R(S_T) = \frac{\sqrt{\hat{q}} \left\lvert \sum_{j=1}^{\hat{q}} \sqrt{n_j}(r'\hat{\beta}_j - \lambda) \right\rvert }{\sum_{j=1}^{\hat{q}} n_j\left(r'\hat{\beta}_j - r'\bar{\beta} \right)^2} \quad , \quad \bar{\beta} = \frac{1}{\hat{q}} \sum_{j=1}^{\hat{q}} \hat{\beta}_j~.
	\end{equation*}    
  \STATE Let $\mathbf{G}$ be the group of $\hat{q} \times 1$ sign changes. For each $g \in \mathbf{G}$, compute $R(gS_T)$ and let their ordered values be:
  \begin{equation*}
  		R^{(1)}(S_T) \leq R^{(2)}(S_T) \leq ... \leq R^{M}(S_T)
  \end{equation*}
  \STATE Let $M$ and $a$ be as defined in equations \ref{eqn--defineM} and \ref{eqn--definea} respectively. Return:
	\begin{align*}
	\phi^{ART}_T(S_T) = \begin{cases}
	1 & \mbox{ if }R(S_T) > R^{(k)}(S_T), \\
	a(S_T) & \mbox{ if }R(S_T) = R^{(k)}(S_T), \\
	0 & \mbox{ otherwise.}
	\end{cases}
	\end{align*}  
\end{algorithmic}
\end{algorithm}

\subsubsection{Inference with Clustered Covariance Estimators}

When there are many clusters, we can simply perform tests based on the usual clustered covariance estimators, using the estimated clusters in place of the unknown true clusters. We briefly review the method below. 

Using the estimated clusters, the clustered covariance estimator is defined as:
\begin{equation}\label{equation--CCE}
	\hat{V}^{CCE} = \frac{1}{\hat{q}}\sum_{j=1}^{\hat{q}} \left(\mathbf{X}_j'\hat{\varepsilon}_j\right)\left(\mathbf{X}_j'\hat{\mathbf{\varepsilon}}_j\right)'
\end{equation}
where $\mathbf{X}_j$ is the $n_j \times p$ matrix formed by stacking row-wise the covariates of all individuals whose estimated cluster is $j$. That is, the $l^{\text{th}}$ row of $\mathbf{X}_j$ is $X_i'$ for some $i$ for whom $g(i) = j$. $\hat{\mathbf{\varepsilon}}_j$ is analogously defined. In other words, $\hat{V}^{CCE}$ is the usual clustered covariance estimator but constructed taking the estimated clusters as true clusters. 

To test the hypothesis in equation \ref{equation--hypothesis}, we form the $t$-statistic:
\begin{equation*}
	T = \frac{\sqrt{\hat{q}}\left(r'\hat{\beta} - \lambda\right)}{\sqrt{r'\hat{V}^{CCE}r}}~.
\end{equation*}
The test is then:
\begin{align}\label{test-cce}
	\phi_T^{CCE} = \begin{cases}
		1 & \mbox{ if } |T| > \Phi^{-1}\left(1-\frac{\alpha}{2}\right) \\
		0 & \mbox{ otherwise.}
	\end{cases}
\end{align}

\subsection{Theoretical Results}\label{section--mainresults}

Our main result shows that clusters are exactly recovered with high probability:

\begin{theorem}[Cluster Recovery]\label{theorem--clusterrecovery}
	Given assumptions \ref{assumption--model}, \ref{assumption--strongmixing}, \ref{assumption--bernstein}, \ref{assumption--uniformity} and \ref{assumption--clusterstructure}, suppose that $L \to \infty$, $L = o(\sqrt{T})$ and $N = O(T^c)$. Then as $T \to \infty$, 
	\begin{equation*}
		P(\hat{g} \cong g) \to 1~.
	\end{equation*}
\end{theorem}
Note that the theorem does not place any lower bound on the $N$. In particular, it is allowed to be fixed. The method also does not place any restrictions on $q$. Our cluster recovery method is therefore consistent with both large and small cluster asymptotics.

Given the ``oracle property" described in the above theorem, it is immediate that approximate randomization tests and CCE-based tests that use the estimated clusters are valid. This is because with high probability, using the estimated clusters is equivalent to using the true clusters. This leads to the following corollaries:

\begin{corollary}[ART-based Test]\label{theorem--sizecontrolfixedq}
		Given assumptions \ref{assumption--model}, \ref{assumption--strongmixing}, \ref{assumption--bernstein}, \ref{assumption--uniformity} and \ref{assumption--clusterstructure}, suppose that $L \to \infty$, $L = o(\sqrt{T})$, $N = O(T^c)$ and that $q$ is fixed as $T \to \infty$. Suppose further that for each $j \in [q]$, there exists $\underline{\lambda} > 0$ so that
		\begin{equation*}
			\lambda_\text{min} \left( \frac{1}{n_j T}\sum_{i \in I_j}^N\sum_{t=1}^T E\left[X_{i,t}X_{i,t}'\right]\right) \geq \underline{\lambda} \quad \mbox{ for all } T \in \mathbb{N}~.
		\end{equation*}
		where $n_j = \lvert\{i : g(i) = j\}\rvert$. Then,
		\begin{equation*}
			\underset{T \to \infty}{\lim \sup} \,\, E\left[\phi^{ART}_T\right] \leq \alpha~.
		\end{equation*}
\end{corollary}

\begin{corollary}[CCE-based Test]\label{theorem--ccesize}
	Given assumptions 1, 2, 3, 4 and 5, suppose that $L \to \infty$,  $L = o(\sqrt{T})$, $N = O(T^c)$ and that $q \to \infty$ is fixed as $T \to \infty$. Suppose further that for some $2 \leq r < \infty$, 
	\begin{equation*}
	T \cdot \frac{\left(\sum_{j = 1}^q |n_j|^r  \right)^{2/r}}{N} \leq C < \infty \quad , \quad \max_{j \in [q]} T \cdot \frac{|n_j|^2}{N} \to 0
	\end{equation*}
	Then,
	\begin{equation*}
		\hat{V}_{CCE}^{-1/2}\left(\hat{\beta} - \beta \right) \overset{d}{\to} \text{N}(0, I)
	\end{equation*}
	where $n_j = \lvert\{i : g(i) = j\}\rvert$ and $\hat{V}_{CCE}$, as displayed in equation \ref{equation--CCE}, is the usual CCE clustered using $\hat{g}$. Furthermore, 
	\begin{equation*}
		\underset{T \to \infty}{\lim \sup} \,\, E\left[\phi_T^{CCE}\right] = \alpha
	\end{equation*}
\end{corollary}

The corollaries above follow from imposing the additional assumptions which are necessary for ART and CCE to yield valid tests. In corollary \ref{theorem--sizecontrolfixedq}, we assume that $q$ is fixed. The additional condition ensures that $\hat{\beta}$ can be estimated cluster-by-cluster. Size control then follows immediately from \cite{crs2017}. 

Similarly, in corollary \ref{theorem--ccesize}, we take $q \to \infty$. The remaining assumptions, taken from \cite{hl2019}, control the amount of heterogeneity across clusters and ensures that CCE with the true clusters yield valid inference. 

%\begin{conjecture}[Size Control with $q \to \infty$]
%	Given assumptions \ref{assumption--model}, \ref{assumption--strongmixing}, \ref{assumption--bernstein}, \ref{assumption--uniformity} and \ref{assumption--clusterstructure}, suppose that $L = o(\sqrt{T})$ and $N = O(T^g)$. Suppose that given $\hat{g}$ we are able to estimate $\hat{\beta}$ cluster-by-cluster. Then the randomization test for the null hypothesis $\hat{\beta} = \beta$ has size that is asymptotically $\alpha$ regardless of our assumptions on $q$. 
%\end{conjecture}

\subsection{Choice of Tuning Parameters}\label{section--crossvalidation}
Our cluster recovery method involves two tuning parameters. We propose to choose them by cross-validation. First, note that $T^{\eta - 1/2} \in [0, 1]$. Equivalently, we search the unit interval for the optimal value of $\tilde{\eta} = T^{\eta - 1/2}$. For $L$, the relevant range over which to search is $\left[0, \sqrt{T}\right]$. Our proposed cross-validation procedure, adapted from that of \cite{bcl2020} and \cite{bl2008}, is as follows. 

First, divide the $T$ time periods into $P = [\log T]$ continuous blocks each of size $[T/\log T] \pm 1$. Number the blocks $1$ to $P$. Fix $L$. For $p \in [P]$, using only the observations in block $p$, compute $\hat{\Sigma}$ as in definition \ref{eqn--sigmahat-def} using windows of size up to $L$. 

Denote this estimate $\hat{\Sigma}_p(L)$. We write $\hat{s}_{p, ij}(L)$ to denote the $(i,j)^{\text{th}}$ entry of $\hat{\Sigma}_p(L)$. For a given $\tilde{\eta}$, let $\tilde{\Sigma}_p(L, \tilde{\eta})$ be $\hat{\Sigma}_p(L)$ thresholded at $\tilde{\eta}$. In other words, 
\begin{align*}
	\tilde{s}_{p,ij}(L, \tilde{\eta}) = \begin{cases}
		\hat{s}_{p,ij}(\tilde{\eta}) & \mbox{ if } \quad \hat{s}_{p, ij}(\tilde{\eta}) \geq \tilde{\eta}\\
		0 & \mbox{ otherwise.}
	\end{cases}
\end{align*}
The cross-validation objective function is:
\begin{equation*}
	\text{CV}(L, \tilde{\eta}) = \sum_{p =1}^P \sum_{p' \neq p} \left\lVert \tilde{\Sigma}_p(L, \tilde{\eta}) - \hat{\Sigma}_{p'}  \right\rVert_F^2
\end{equation*}
The cross-validation value of $L$ and $\tilde{\eta}$ are then:
\begin{equation*}
	\left(L^*, \eta^*\right) = \underset{\tilde{\eta} \in [0, 1]\, , \, L \in \left[1, \sqrt{T}\right]}{\arg \min} \,\, \text{CV}\left(L, \tilde{\eta}\right)
\end{equation*}
Intuitively, our cross-validation procedure relies on stationarity. If the correlation between units are stable over time, the properly thresholded estimator for some block $p$ should be close to the unthresholded estimators obtained from all other blocks. 

Although the theoretical results in section \ref{section--mainresults} do not allow for data-dependent choices of $\eta$, the simulations in section \ref{section--simulations} suggest that cross-validation works well in practice. 

\begin{remark}
	In the objective function above, we cannot use the thresholded estimator for both ${p}$ and $p'$. Such an objective function would be minimized by $\tilde{\eta} = 1$, since eliminating all entries other than the diagonal will lead to a cross-validation error of $0$. 
\end{remark}

\section{Relation to Existing Literature}\label{section--existinglit}

In this subsection, we discuss the work most closely related to ours \cite{bcl2020} (section \ref{subsubsection--bcl}), as well as the literatures on tests for level of clustering (section \ref{subsubsection--levelofclustering}) and panel data with interactive fixed effects (section \ref{subsubsection--interactivefes}). Finally, we discuss how our work can be useful even when researchers take the ``design-based" approach of \cite{aaiw17} (section \ref{section--aaiw}). 

\subsection{\cite{bcl2020}}\label{subsubsection--bcl}

Our paper is most closely related to \cite{bcl2020} (henceforth BCL), which to our knowledge is the only other paper that is explicitly concerned with inference in settings with unknown clusters. The starting point of their method is the observation that dependence across units affect the variance $\hat{\beta}$ only through:
\begin{align*}
V & = \text{Var}\left(\frac{1}{\sqrt{NT}} \sum_{i=1}^N \sum_{t = 1}^T X_{i,t}U_{i,t} \right) \\
& = \frac{1}{NT}\sum_{t=1} E\left[\mathbf{X}_t U_t U_t'\mathbf{X}_t  \right] + \frac{1}{NT}\sum_{h=1}^{T-1}\sum_{t=h+1}^T \left( E\left[\mathbf{X}_t U_t U_{t-h}'\mathbf{X}_{t-h}  \right] + E\left[\mathbf{X}_{t-h} U_{t-h} U_t'\mathbf{X}_t  \right]\right) \\
& = \frac{1}{N} \sum_{i,j} V_{i,j}~,
\end{align*}
where
\begin{equation*}
V_{i,j} = \frac{1}{T}\sum_{t=1}^T E\left[{X}_{i,t} U_{i,t} U_{j,t}{X}_{j,t}'  \right] + \frac{1}{T}\sum_{h=1}^{T-1}\sum_{t=h+1}^T \left( E\left[{X}_{i,t} U_{i,t} U_{j,t-h}{X}_{j,t-h}'  \right] + E\left[{X}_{i,t-h} U_{i,t-h} U_{j,t}{X}_{j,t}'  \right]\right)~.
\end{equation*}
Consider the Newey-West estimator of $V_{i,j}$:
\begin{equation*}
	\hat{V}_{i,j} = \frac{1}{T} \sum_{t=1}^T X_{i,t}\hat{U}_{i,t}\hat{U}_{i,t}'X_{i,t} + \frac{1}{T} \sum_{h = 1}^L \omega(h, L) \sum_{t = h+1}^L \left(X_{i,t}\hat{U}_{i,t} \hat{U}_{j.t-h}X'_{j, t-h}  + X_{i,t-h}\hat{U}_{i,t-h} \hat{U}_{j,t}X'_{j, t} \right)
\end{equation*}
Since they are consistent for $V_{i,j}$, it seems reasonable to construct the estimator:
\begin{equation*}
	\hat{V} = \frac{1}{N} \sum_{i,j} \hat{V}_{i,j}~.
\end{equation*}
However, it turns out that when $N$ is large, such an estimator ``accumulates a large number of cross-sectional estimation noises[sic]" (BCL, pg 3). Conventional clustered covariance estimator overcomes this problem by setting $V_{i,j}$ for which $i,j$ are in different clusters to $0$, so that a large number of terms do not need to be estimated. 

In the absence of this information, BCL assumes that the set of $V_{i,j}$'s are sparse -- that is, that they are mostly $0$ -- and propose to use a thresholding method to identify the entries which are $0$. In effect, they employ the estimator:
\begin{equation*}
	\hat{V}^{BCL} = \frac{1}{N} \sum_{i,j} \hat{V}_{i,j}\mathbf{1}\left\{ \hat{V}_{i,j} \geq \lambda_{i,j}    \right\}
\end{equation*}
which can then be used to construct an estimator for the variance-covariance matrix of $\hat{\beta} - \beta$. The authors show that their estimator is consistent and leads to tests which are valid. 

Our method is similar to BCL since we also threshold an long-run correlation matrix. In fact, $\hat{\sigma}_{i,j} = \iota' \hat{V}_{i,j}\iota$. The advantage of BCL over our method is that they do not require the existence of clusters for inference. In particular, they could accommodate a dependency structure in which for any two units, one can find a chain of units along which pairwise correlation is not $0$. Our method would have poor power in such a situation since there is only one cluster. 

However, in a cluster context, BCL's sparsity assumption implies $q \to \infty$. Using ART with the recovered clusters, we are able to accommodate a fixed $q$ setting. This means, for example, that we allow a given unit to be correlated with $O(N)$ other units, a setting for which BCL is unsuitable. 
In addition, we state a cluster recovery result which is novel. This is not found in BCL, although this is because they do not assume the existence of a cluster-structure. 

Furthermore, we relax two potentially restrictive assumptions in BCL. First, unlike BCL, we do not require $N \to \infty$. Note that both methods require $T \to \infty$ and $N = O(T^c)$ for some $c$.\footnote{In fact, BCL allows $N$ to grow at an exponential rate in relation to $T$. Our method does as well, though for ease of exposition, we have opted to state our result in terms of arbitrarily large polynomial rate growth.} Secondly, we do not impose any lower bound on the $\alpha$-mixing coefficients across time. Such a lower bound excludes the setting in which observations are independent over time. This is unnatural since we expect that the absence of dependence would make the inference problem easier. 

\subsection{Tests for Level of Clustering}\label{subsubsection--levelofclustering}

Our concern with the fact that clusters are typically unknown ex ante is shared by the literature on tests for level of clustering (\citealt{im2016}, \citealt{mnw2020} and \citealt{cai--testforclustering}). These tests assume that researchers are able to identify clusters which are independent and use this information to test whether a conjectured finer level of clustering is valid. 

Researchers could plausibly use a sequence of these tests to discover the level of clustering by testing the validity of increasingly fine sub-clusters. However, such a method is unsuitable for use in discovering clusters. First, these tests would have to be adjusted for multiple testing, especially if the goal is inference. No sophisticated method of adjustment is available and Bonferroni corrections will lead to tests which are too conservative. Secondly, these methods require the conjectured clustering to be nested within the valid clustering that is known. This is restrictive. 

Lastly and most importantly, researchers may not know of any independent clusters. With these methods, assuming that there is one cluster containing all the observations is not an option. This is because \cite{im2016} and \cite{mnw2020} require at least two clusters to be computable, while \cite{cai--testforclustering} has trivial power when there is only one cluster. 

Our method and BCL therefore complement this literature by providing methods that choose the level of clustering directly, without requiring input from the researcher beyond specifying the linear model. 

\subsection{Panel Regression with Interactive Fixed Effects}\label{subsubsection--interactivefes}

Our paper is also related to a large literature on panel regression models with interactive fixed effects (see for instance \citealt{bai2009}, \citealt{mw2015}, \citealt{bm2015} among many more.) Relative to our model, these papers further assume that the error term has a factor structure:
\begin{align*}
	U_{i,t} = \sum_{r = 1}^{R_0} \lambda_{i,r} f_{t,r} + E_{i,t}~.
\end{align*}
In the above equation, $E_{i,t}$ is white noise, although $f_{t,r}$ maybe correlated with $X_{i,t}$. If $f_{t,r}$ are exogenous to $X_{i,t}$, the above model would be nested in ours, which accommodates richer patterns of correlation than the factor structure. 

Allowing $f_{t,r}$ to be endogenous necessitates estimation procedures that are either computationally difficult, or require further assumptions on the $X_{i,t}$'s. Our procedure is unable to accommodate endogenous $f_{t,r}$, but it is computationally simple without these additional requirements. 

\subsection{Design-Based Approach}\label{section--aaiw}

\cite{aaiw17} (henceforth AAIW) advocate a ``design-based" perspective on the issue of clustering. They are concerned with two types of design issues. Clustering is a sampling design issue when samples are drawn according to a two-stage process, in which the first step entails sampling clusters and individuals are drawn only in the second stage. Because clusters could differ systematically and not all clusters are observed, researchers wanting to learn the average effect over all clusters have to take into account the uncertainty introduced by this clustered sampling procedure. Conversely, clustering is an experimental design issue when clusters of units, rather than individual ones, are assigned to a treatment by the assignment mechanism. Within each cluster, we observe only one of two potential outcomes. Because clusters could differ systematically in their potential outcomes, researchers who want to learn an average effect must again cluster their standard errors. AAIW then call on researcher to be explicit in their beliefs about the sampling and treatment assignment process, and use these beliefs to guide their clustering decisions.

For a concrete example, suppose we are interested in studying the returns to education in the US. In other words, we are interested in the average effect of an additional year of education on income for the average individual in the US. Suppose we only observe individuals in 25 randomly chosen states, but that education is randomly assigned to all individual within this sample. By AAIW, the researcher should cluster at the state level due to clustering in the sampling process. Suppose instead that we observe individuals in all states, but that all individuals in a given state were randomly assigned the same years of education. Then, because of clustering in experimental design, standard errors should again be clustered at the state level. Finally, suppose that we observe all individuals, and that years of education was randomly assigned to each individual. Then there is no clustering arising from sample design or experimental design. There is no need to cluster standard errors, even though individuals in the same state may have scores $X_{i,t}U_{i,t}$ that are correlated (provided of course that these unobservables are not correlated with years of education). As such, researchers should cluster their standard errors at the state only if they believe that their sample is drawn from a subset of states, or if they believe that education was randomly assigned at the state level. 

At first sight, the design-based approach contradicts the conventional model-based approach to clustering: the latter demands that researchers cluster their standard errors when there is correlation in the scores of the clusters. This contradiction can be resolved once we realise that the design-based approach is concerned with a different estimand: the conditional treatment effect. Appendix \ref{appendix--aaiw} provides a simple illustration of this point. 

The approach of AAIW is theoretically insightful, but not necessarily helpful for a researcher deciding how to cluster. This is because they require researchers to know the sampling or treatment assignment process, which researchers may not know. Our procedure can still be useful in such a scenario. For example, suppose the treatment variable, $D_{i,t}$, is $\alpha$-mixing over time. Then, applying our cluster recovery method with $D_{i,t}$ replacing $X_{i,t}\hat{U}_{i,t}$ will yield clusters across which treatment assignment is asymptotically independent. More generally, applying our method in their framework will still yield clusters that are valid -- that is, clusters with observables and unobservables that are asymptotically independent. The trade-off is that these clusters might be too coarse, leading to relatively lower power. 

Hence, our method is compatible with the design-based approach to clustering. Despite the potential power loss, researchers might find our data-driven method useful, especially when they have little information or are unwilling to make assumptions on the true sampling and treatment assignment mechanisms.

\section{Monte Carlo Simulations}\label{section--simulations}

In this section we study the finite sample performance of our method via Monte Carlo simulations. Section \ref{section--simulations--clusterrecover}, evaluates the method's ability to reliably recover clusters from the data. Section \ref{section--simulations--test} considers the size and power of tests based on the recovered clusters. 

We employ the following data generating process:
\begin{align*}
	Y_{it} & = \beta + \frac{1}{2}V_{g(i)t} + \frac{1}{2}U_{it}  \\
	U_{it} & = \rho U_{it-1} + \varepsilon_{it} \quad , \quad \varepsilon_{it} \sim \text{N}(0, 1) \\
	V_{jt} & = \phi V_{jt-1} + \nu_{jt} \quad , \quad \nu_{jt} \sim \text{N}(0, 1)
\end{align*}
where $g(i)$ denotes the cluster to which each unit $i$ belongs. Specifically, we set $\rho = \phi = 0.2$ and $\beta = 1$. We study the performance of the test as the following parameters vary: $q \in \{5, 10, 25\}$, $n \in \{50, 200\}$, $t \in \{100, 200\}$. We select the tuning parameters $\eta$ and $L$ by the cross-validation procedure of section \ref{section--crossvalidation}. Our results show that the method is effective in recovering the true clusters and leads to tests which are both valid and powerful. 

\subsection{Cluster Recovery}\label{section--simulations--clusterrecover}

Table \ref{tab--clusterrecovery} presents results on cluster recovery. We assess the quality of the estimated clusters on purity: the largest number of individuals in each estimated cluster that truly belong to the same cluster. We consider both the minimum purity across the estimated clusters as well as average purity. Our second criterion is the number of estimated clusters $\hat{q}$. The method achieves perfect recovery if and only if minimum and average cluster purity are both equal to $1$ and $\hat{q} = q$. 

We see that the method reliably recovers clusters in the data. Both minimum and average purity of the estimated clusters are high. For $q = 5$, $\hat{q}$ is close to $q$, suggesting that the clusters are very high quality. For $q = 10$ and $q = 25$, we still have high cluster purity. However, the $\hat{q}$ can be very large when $n =100$. This means that the estimated clusters are splintered version of the true clusters. By the time $n = 200$, however, the problem mostly goes away. $q = 25$ is the most challenging for cluster recovery. This is unsurprising since with $25$ clusters, the long-run covariance matrix is mostly noise, making cluster recovery a challenging task. Nonetheless, it performs reasonably well. 

% Table generated by Excel2LaTeX from sheet 'clusterpurity CV paper version'
\begin{table}[h!]
  \centering \small 
    \begin{tabular}{>{\centering}p{.75cm}>{\centering}p{.75cm}>{\centering}p{.75cm}ccc}
        \cmidrule{1-6}\morecmidrules\cmidrule{1-6}
        \multirow{2}[2]{*}{$q$} & \multirow{2}[2]{*}{$N$} & \multirow{2}[2]{*}{$T$} & Min.  & Avg.  & \multirow{2}[2]{*}{Avg. $\hat{q}$} \\
            &       &       & Pur   & Pur   &  \\
	    \midrule
	    \multirow{4}[4]{*}{5} & \multirow{2}[2]{*}{50} & 100   & 0.890 & 0.962 & 4.924 \\
		          &       & 200   & 0.980 & 0.995 & 4.956 \\
		\cmidrule{2-6}          & \multirow{2}[2]{*}{200} & 100   & 0.542 & 0.730 & 3.319 \\
		          &       & 200   & 0.847 & 0.944 & 4.598 \\
		\midrule
		\multirow{4}[4]{*}{10} & \multirow{2}[2]{*}{50} & 100   & 1.000 & 1.000 & 25.133 \\
		          &       & 200   & 1.000 & 1.000 & 10.130 \\
		\cmidrule{2-6}          & \multirow{2}[2]{*}{200} & 100   & 0.998 & 1.000 & 31.775 \\
		          &       & 200   & 0.997 & 1.000 & 9.994 \\
		\midrule
		\multirow{4}[3]{*}{25} & \multirow{2}[2]{*}{50} & 100   & 1.000 & 1.000 & 48.580 \\
		          &       & 200   & 1.000 & 1.000 & 38.264 \\
		\cmidrule{2-6}          & \multirow{2}[1]{*}{200} & 100   & 1.000 & 1.000 & 198.279 \\
		          &       & 200   & 1.000 & 1.000 & 35.782 \\
	    \cmidrule{1-6}\morecmidrules\cmidrule{1-6}          
    \end{tabular}%
  \caption{Results for cluster recovery. Minimum purity is the purity of the worst cluster. Average purity is the average purity of all clusters. Average $\hat{q}$ is the average number of clusters  predicted by the algorithm. Results based on 1,000 simulations.}
  \label{tab--clusterrecovery}%
\end{table}%

\subsection{Test Performance}\label{section--simulations--test}

Table \ref{tab--size} presents type I error of our test as well as that of \cite{bcl2020} for $\alpha = 10\%$. We see that the ART version of our test has good size control when $q = 5$. When $q = 10$, the test initially over-rejects when $t = 100$. However, type I error is close to $10\%$ once $t = 200$. For $q = 5, 10$, the test performs very similarly to the oracle version of ART. This is unsurprising since the clusters are very well estimated at these values. When $q = 25$, ART is over-rejects at close to 20\%. The fact that clusters are relatively poorly estimated affects size control here, though we see that as $T$ grows, the problem quickly goes away. In contrast, the CCE version of our method does well when $q = 25$. It has poor size control when $q = 5$, due to the well known fact that CCE's are downward biased when $q$ is small. However, as $q$ grows, the CCE method controls size well. In particular, it handles the situation with $q = 25$ better than the ART method. This suggests that the CCE method is less sensitive to clusters that are wrongly estimated. In general, our simulation results concord with the theoretical results in section \ref{section--mainresults}. Turning to BCL, we see that across all parameter values, the method is fairly conservative, with size below 1\%. 

% Table generated by Excel2LaTeX from sheet 'inference paper version'
\begin{table}[h!]
  \centering \small 
    \begin{tabular}{>{\centering}p{.75cm}>{\centering}p{.75cm}>{\centering}p{.75cm}ccccccc}
    		\cmidrule{1-10}\morecmidrules\cmidrule{1-10}    
\cmidrule{4-5}\cmidrule{7-8}\cmidrule{10-10}          &       &       & \multicolumn{2}{c}{Oracle} &       & \multicolumn{2}{c}{Our Method} &       & \multirow{2}[4]{*}{BCL} \\
\cmidrule{4-5}\cmidrule{7-8}    $q$   & $N$   & $T$   & ART   & CCE   &       & ART   & CCE   &       &  \\
    \midrule
    \multirow{4}[4]{*}{5} & \multirow{2}[2]{*}{50} & 100   & 0.053 & 0.181 &       & 0.053 & 0.199 &       & 0.007 \\
          &       & 200   & 0.076 & 0.184 &       & 0.074 & 0.187 &       & 0.009 \\
\cmidrule{2-5}\cmidrule{7-8}\cmidrule{10-10}          & \multirow{2}[2]{*}{200} & 100   & 0.073 & 0.197 &       & 0.032 & 0.265 &       & 0.008 \\
          &       & 200   & 0.076 & 0.197 &       & 0.053 & 0.212 &       & 0.005 \\
\cmidrule{1-5}\cmidrule{7-8}\cmidrule{10-10}    \multirow{4}[4]{*}{10} & \multirow{2}[2]{*}{50} & 100   & 0.094 & 0.134 &       & 0.237 & 0.207 &       & 0.005 \\
          &       & 200   & 0.110 & 0.139 &       & 0.114 & 0.140 &       & 0.006 \\
\cmidrule{2-5}\cmidrule{7-8}\cmidrule{10-10}          & \multirow{2}[2]{*}{200} & 100   & 0.113 & 0.147 &       & 0.310 & 0.187 &       & 0.010 \\
          &       & 200   & 0.100 & 0.141 &       & 0.100 & 0.140 &       & 0.007 \\
\cmidrule{1-5}\cmidrule{7-8}\cmidrule{10-10}    \multirow{4}[3]{*}{25} & \multirow{2}[2]{*}{50} & 100   & 0.096 & 0.111 &       & 0.176 & 0.187 &       & 0.010 \\
          &       & 200   & 0.099 & 0.116 &       & 0.153 & 0.145 &       & 0.001 \\
\cmidrule{2-5}\cmidrule{7-8}\cmidrule{10-10}          & \multirow{2}[1]{*}{200} & 100   & 0.105 & 0.114 &       & 0.453 & 0.442 &       & 0.009 \\
          &       & 200   & 0.100 & 0.119 &       & 0.206 & 0.132 &       & 0.002 \\
    		\cmidrule{1-10}\morecmidrules\cmidrule{1-10}              
    \end{tabular}%
  \caption{Rejection rate for $\alpha = 10\%$ under the null hypothesis $\beta_0 = 1$. ART is approximate randomization test. CCE is the $t$-test based on cluster-robust covariance estimator. MST refers to the clusters recovered using our method. BCL refers to the thresholded CCE of \cite{bcl2020}. ``Oracle" refers to running ART and CCE on the true clusters.}
  \label{tab--size}%
  
      \begin{tabular}{>{\centering}p{.75cm}>{\centering}p{.75cm}>{\centering}p{.75cm}ccccccc}
  	\cmidrule{1-10}\morecmidrules\cmidrule{1-10}              
            &       &       & \multicolumn{2}{c}{Oracle} &       & \multicolumn{2}{c}{Our Method} &       & \multirow{2}[3]{*}{BCL} \\
  \cmidrule{4-5}\cmidrule{7-8}    $q$   & $N$   & $T$   & ART   & CCE   &       & ART   & CCE   &       &  \\
      \midrule
      \multirow{4}[4]{*}{5} & \multirow{2}[2]{*}{50} & 100   & 0.172 & 0.369 &       & 0.147 & 0.390 &       & 0.069 \\
            &       & 200   & 0.301 & 0.541 &       & 0.287 & 0.540 &       & 0.096 \\
  \cmidrule{2-5}\cmidrule{7-8}\cmidrule{10-10}          & \multirow{2}[2]{*}{200} & 100   & 0.176 & 0.405 &       & 0.072 & 0.456 &       & 0.097 \\
            &       & 200   & 0.302 & 0.558 &       & 0.228 & 0.572 &       & 0.127 \\
  \cmidrule{1-5}\cmidrule{7-8}\cmidrule{10-10}    \multirow{4}[4]{*}{10} & \multirow{2}[2]{*}{50} & 100   & 0.447 & 0.507 &       & 0.628 & 0.614 &       & 0.168 \\
            &       & 200   & 0.662 & 0.728 &       & 0.668 & 0.730 &       & 0.238 \\
  \cmidrule{2-5}\cmidrule{7-8}\cmidrule{10-10}          & \multirow{2}[2]{*}{200} & 100   & 0.510 & 0.568 &       & 0.662 & 0.616 &       & 0.206 \\
            &       & 200   & 0.733 & 0.783 &       & 0.733 & 0.783 &       & 0.291 \\
  \cmidrule{1-5}\cmidrule{7-8}\cmidrule{10-10}    \multirow{4}[3]{*}{25} & \multirow{2}[2]{*}{50} & 100   & 0.713 & 0.733 &       & 0.820 & 0.822 &       & 0.348 \\
            &       & 200   & 0.939 & 0.948 &       & 0.961 & 0.967 &       & 0.611 \\
  \cmidrule{2-5}\cmidrule{7-8}\cmidrule{10-10}          & \multirow{2}[1]{*}{200} & 100   & 0.832 & 0.844 &       & 0.965 & 0.966 &       & 0.504 \\
            &       & 200   & 0.975 & 0.980 &       & 0.961 & 0.982 &       & 0.780 \\
  	\cmidrule{1-10}\morecmidrules\cmidrule{1-10}                        
      \end{tabular}%
  	\caption{Rejection rate for $\alpha = 10\%$ under the alternative hypothesis $\beta_0 = 0.95$. ART is approximate randomization test. CCE is the $t$-test based on cluster-robust covariance estimator. MST refers to the clusters recovered using our method. BCL refers to the thresholded CCE of \cite{bcl2020}. ``Oracle" refers to running ART and CCE on the true clusters.}    
  	\label{tab--power95}%
  
\end{table}%

Table \ref{tab--power95} presents the power of our test as well as that of \cite{bcl2020} for $\alpha = 10\%$ when $\beta_0 = 0.95$. Across the parameter values considered, both the ART and CCE version of our procedure has least 10\% more power than BCL, often much more. The difference is less stark once we turn to table \ref{tab--power90}. Here, the null hypothesis is so clearly wrong that both methods reject it at very high rates. 

% Table generated by Excel2LaTeX from sheet 'inference paper version'
\begin{table}[htbp]
  \centering \small
    \begin{tabular}{>{\centering}p{.75cm}>{\centering}p{.75cm}>{\centering}p{.75cm}ccccccc}
	\cmidrule{1-10}\morecmidrules\cmidrule{1-10}                  
          &       &       & \multicolumn{2}{c}{Oracle} &       & \multicolumn{2}{c}{Our Method} &       & \multirow{2}[3]{*}{BCL} \\
\cmidrule{4-5}\cmidrule{7-8}    $q$   & $N$   & $Tt$   & ART   & CCE   &       & ART   & CCE   &       &  \\
    \midrule
    \multirow{4}[4]{*}{5} & \multirow{2}[2]{*}{50} & 100   & 0.447 & 0.757 &       & 0.387 & 0.769 &       & 0.400 \\
          &       & 200   & 0.715 & 0.935 &       & 0.688 & 0.936 &       & 0.661 \\
\cmidrule{2-5}\cmidrule{7-8}\cmidrule{10-10}          & \multirow{2}[2]{*}{200} & 100   & 0.501 & 0.808 &       & 0.183 & 0.822 &       & 0.465 \\
          &       & 200   & 0.725 & 0.952 &       & 0.534 & 0.949 &       & 0.695 \\
\cmidrule{1-5}\cmidrule{7-8}\cmidrule{10-10}    \multirow{4}[4]{*}{10} & \multirow{2}[2]{*}{50} & 100   & 0.911 & 0.938 &       & 0.954 & 0.961 &       & 0.730 \\
          &       & 200   & 0.991 & 0.994 &       & 0.990 & 0.994 &       & 0.949 \\
\cmidrule{2-5}\cmidrule{7-8}\cmidrule{10-10}          & \multirow{2}[2]{*}{200} & 100   & 0.943 & 0.960 &       & 0.962 & 0.970 &       & 0.800 \\
          &       & 200   & 0.998 & 0.999 &       & 0.998 & 0.999 &       & 0.968 \\
\cmidrule{1-5}\cmidrule{7-8}\cmidrule{10-10}    \multirow{4}[3]{*}{25} & \multirow{2}[2]{*}{50} & 100   & 0.999 & 0.999 &       & 1.000 & 1.000 &       & 0.976 \\
          &       & 200   & 1.000 & 1.000 &       & 1.000 & 1.000 &       & 1.000 \\
\cmidrule{2-5}\cmidrule{7-8}\cmidrule{10-10}          & \multirow{2}[1]{*}{200} & 100   & 1.000 & 1.000 &       & 1.000 & 1.000 &       & 0.994 \\
          &       & 200   & 1.000 & 1.000 &       & 1.000 & 1.000 &       & 1.000 \\
	\cmidrule{1-10}\morecmidrules\cmidrule{1-10}                        
    \end{tabular}%
	\caption{Rejection rate for $\alpha = 10\%$ under the alternative hypothesis $\beta_0 = 0.9$. ART is approximate randomization test. CCE is the $t$-test based on cluster-robust covariance estimator. MST refers to the clusters recovered using our method. BCL refers to the thresholded CCE of \cite{bcl2020}. ``Oracle" refers to running ART and CCE on the true clusters.}    
	\label{tab--power90}%
\end{table}%

In summary, our simulation result suggests that our cluster recovery method finds situations with large $q$ more challenging, but generally performs well. Furthermore, inference method that combined our recovered clusters with either ART or CCE are able to control size well, and has much more power than BCL. Our method is hence useful for inference in panel data with unknown clusters. 

%\section{Application}\label{section--application}

\section{Conclusion}\label{section--conclusion}

We provide a method for inference in panel data with unknown clusters. We propose a procedure to help researchers discover clusters in panel data. Our method is based on thresholding an estimated long-run variance-covariance matrix and requires the panel to be large in the time dimension, but imposes no lower bound on the number of units. We provide a novel theoretical result showing exact cluster recovery with high probability. Furthermore, the recovered clusters can be combined with either approximate randomization tests or tests based on clustered covariance estimators to yield valid inference. The test based on approximate randomization test controls size even when the number of clusters is small, a setting that is not currently handled by existing papers. Simulation results show that our method has more power than existing methods, making it a useful addition to the toolbox of applied economists. 

\clearpage
\newpage 

\bibliographystyle{chicago}
\bibliography{clusters}

\clearpage

\appendix

\section*{Appendices}

\section{Proofs}\label{appendix--proofs}

\subsection{Proof of Theorem \ref{theorem--clusterrecovery}}

Our proof for theorem \ref{theorem--clusterrecovery} is divided into the following lemmata:

%\begin{definition}[Spurious Links]
%	We say that the tree $\hat{G}$ has a spurious link $(i,j)$ if $(i,j) \in E(\hat{G})$ but $\sigma_{i,j} = 0$. 
%\end{definition}
%
%\begin{lemma}[Sufficiency of Separation]\label{lemma--algorithmcorrectness}
%	Suppose that for all $(i,j)$ such that $\sigma_{i,j} \neq 0$ and all $(i', j')$ such that $\sigma_{i',j'} = 0$, we have that $\hat{\sigma}_{i,j} > \hat{\sigma}_{i',j'}$. Then the empirical maximum weight spanning tree, $\hat{G}$, comprises $q$ trees, each corresponding to a true cluster $j \in [q]$, joined together by $q-1$ spurious links. 
%\end{lemma}

\begin{lemma}[Convergence of $\hat{\beta}$]\label{lemma--convergenceofbetahat}
	Given assumptions \ref{assumption--model}, \ref{assumption--strongmixing}, \ref{assumption--bernstein} and \ref{assumption--uniformity}, 
	\begin{align*}
		& P\left( \left\lVert \hat{\beta} - \beta \right\rVert > \varepsilon \right) \\
		& \quad \leq N \cdot O\left(\sqrt{T-L} \exp\left( -C_3(T-L)\varepsilon^2 \right)  \right) + N \cdot O \left(\frac{(T-L)}{\varepsilon} \exp\left( -C_4({T-L})^{\kappa/2} \right)  \right) \\ 
		& \quad + N\frac{p(p-1)}{2} O\left(\sqrt{T} \exp\left( -C_7T \right)  \right) + N\frac{p(p-1)}{2} O \left(T\exp\left( -C_8{T}^{\kappa/2} \right)  \right)~.
	\end{align*}
\end{lemma}

\begin{definition}
	Define 	
	\begin{align*}
		\tilde{\sigma}_{i,j}^{a,b} = E[X_{i,t}^{(a)}U_{i,t}X_{j,t}^{(b)}U_{j,t}] + \sum_{h=1}^{L} \omega(h, L) \left(	E[X_{i,t}^{(a)}{U}_{i,t}X_{j,t-h}^{(b)}{U}_{j,t-h}] + E[X_{i,t-h}^{(a)}{U}_{i,t-h}X_{j,t}^{(b)}{U}_{j,t}] \right) ~.
	\end{align*}
	where $\omega(h, L)$ is the Bartlett kernel (see definition \ref{definition--Bartlett}). Further define:
	\begin{equation*}
		\tilde{\sigma}_{i,j} = \sum_{a = 1}^p \sum_{b = 1}^p \left\lvert \tilde{\sigma}_{i,j}^{a,b} \right\rvert
	\end{equation*}	
\end{definition}

\begin{lemma}[Closeness of $\tilde{\sigma}_{i,j}$ and $\sigma_{i,j}$]\label{lemma--closenesstilde}
Given assumptions \ref{assumption--strongmixing} and \ref{assumption--uniformity},
\begin{equation*}
	\max_{i,j \in [N]} |\tilde{\sigma}_{i,j} - {\sigma}_{i,j}| \to 0 \mbox{ as } L \to \infty~. 
\end{equation*}
\end{lemma}

\begin{lemma}[Concentration of $\hat{\sigma}_{i,j}$]\label{lemma--closenesshat}
	%where $w(h, L)$ is the Bartlett kernel with bandwidth $L$ and horizon $h$. 
	Given assumptions 1, 3, 4 and 5, suppose $L = o(\sqrt{T})$ and $N = O(T^g)$ for some $g < \infty$. Then, for all $0 < \eta < 1/2$,
	\begin{equation*}
		P\left(\max_{i,j \in [N]} |\hat{\sigma}_{i,j} - \tilde{\sigma}_{i,j}| > T^{\eta -1/2} \right) \to 0
	\end{equation*}
\end{lemma}

\subsubsection{Main Proof}

%\begin{proof}
	In the following, $\lVert \cdot \rVert$ is the $L^2$ norm for a vector and the operator norm for a matrix. Then we can write:
	\begin{equation*}
		\max_{i,j \in [N]} |\hat{\sigma}_{i,j} - \sigma_{i,j}| \leq \max_{i,j \in [N]} |\hat{\sigma}_{i,j} - \tilde{\sigma}_{i,j}| + \max_{i,j \in [N]} |\tilde{\sigma}_{i,j} - {\sigma}_{i,j}|~.
	\end{equation*} 
	By Lemma \ref{lemma--closenesstilde}, we have that the second term is smaller than $\underline{\sigma}/4$ for $T$ large enough. By Lemma \ref{lemma--closenesshat}, we have that 
	\begin{equation*}
		P\left(\max_{i,j \in [N]} |\hat{\sigma}_{i,j} - \tilde{\sigma}_{i,j}| >  \frac{\underline{\sigma}}{4} \right) \to 0
	\end{equation*}
	As such, for all $i,j$ such that $\sigma_{i,j} \neq 0$, we have that with probability approaching 1,
	\begin{equation*}
		P\left(\max_{i,j \in [N]} |\hat{\sigma}_{i,j} - {\sigma}_{i,j}| > \frac{\underline{\sigma}}{2}\right) \to 0~.
	\end{equation*}	
	Hence, with probability approaching 1, we have that for $T$ large enough, 
	\begin{equation*}
		\frac{\hat{\sigma}_{i,j} }{\sqrt{ \hat{\sigma}_{i,i} \hat{\sigma}_{j,j}}} \geq \frac{\underline{\sigma} - \frac{2 \log_2(T)}{T^{\eta}}}{\sqrt{\overline{\sigma}^2 - 2 \underline{\sigma} \frac{2 \log_2(T)}{T^{\eta}} + \left(\frac{2 \log_2(T)}{T^{\eta}} \right)^2  }} \geq \frac{\underline{\sigma}}{2 \overline{\sigma}} \,\, \mbox{for all } i, j~.
	\end{equation*}
	Meanwhile, if $i,j$ is such that $\sigma_{i,j} = 0$, then $\tilde{\sigma}_{i,j} - \sigma_{i,j} = 0$. Thus we have that 
	\begin{equation*}
		P\left(\max_{i,j \in [N]} |\hat{\sigma}_{i,j} - {\sigma}_{i,j}| > T^{\eta -1/2} \right) \to 0 \Leftrightarrow \max_{\sigma_{i,j} = 0} |\hat{\sigma}_{i,j}| \leq T^{\eta - 1/2} \,\, \mbox{w.p.a.} 1
	\end{equation*}	
	Therefore, 
	\begin{equation*}
		\frac{\hat{\sigma}_{i,j} }{\sqrt{ \hat{\sigma}_{i,i} \hat{\sigma}_{j,j}}} \leq \frac{T^{\eta -1/2}}{\sqrt{\overline{\sigma}^2 - 2 \underline{\sigma} \frac{2 \log_2(T)}{T^{\eta}} + \left(\frac{2 \log_2(T)}{T^{\eta}} \right)^2  }} \leq T^{\tilde{\eta} -1/2} \,\, \mbox{for $T$ large enough,}
	\end{equation*}
	where $0 < \tilde{\eta} < 1/2$. Hence, with probability approaching 1, our strategy of deleting links if and only if $\frac{\hat{\sigma}_{i,j} }{\sqrt{ \hat{\sigma}_{i,i} \hat{\sigma}_{j,j}}} \leq T^{\eta - 1/2}$ decides that two individuals are independent if and only if they indeed are independent.
	
	Putting our results together, we have that $P(\hat{g} \cong g) \to 1$ as $T \to \infty$ provided that $N = O\left(T^g\right)$ and $L = o\left(\sqrt{T}\right)$, $0 < \eta < 1/2$. 

\subsubsection{Proof of Lemma \ref{lemma--convergenceofbetahat}}

%\begin{proof}
	Write:
	\begin{equation*}
		\lVert \hat{\beta} - \beta \rVert \leq \left\lVert \left(\frac{1}{NT}\sum_{i=1}^N\sum_{t=1}^T X_{it}X_{it}'\right)^{-1}\right\rVert \cdot  \left\lVert  \frac{1}{NT}\sum_{i=1}^N\sum_{t=1}^T X_{it}U_{it}	\right\rVert
	\end{equation*}
	By the union bound, as well as $N$ applications of the Theorem 1.4 in Bosq (1998),
	\begin{align*}
		& P\left( \left\lvert \frac{1}{NT}\sum_{i=1}^N\sum_{t=1}^T X^{(l)}_{it}U_{it} \right\rvert > \varepsilon \right) 
		\leq \sum_{i=1}^N P\left( \left\lvert \frac{1}{NT} \sum_{t=1}^T X^{(l)}_{it}U_{it} \right\rvert > \frac{\varepsilon}{N} \right) \\
		& \leq N \cdot O\left(\sqrt{T-L} \exp\left( -C_3(T-L)\varepsilon^2 \right)  \right) + N \cdot O \left(\frac{(T-L)}{\varepsilon} \exp\left( -C_4({T-L})^{\kappa/2} \right)  \right)
	\end{align*}
	where the bound is exactly as in equation (\ref{eqn--to0bound}), except for the factor of $N$ that arose from the union bound.
		
	By continuity of the matrix inverse at $\frac{1}{NT}\sum_{i=1}^N\sum_{t=1}^T E[X_{it}X_{it}']$ as well as the continuity of $\lVert \cdot \rVert$, we can find an $\bar{\varepsilon}$ such that
	\begin{align*}
		\left\lVert\frac{1}{T} \sum_{t=h+1}^T {X}^{(l)}_{i,t}{X}^{(m)}_{j,t} - E\left[{X}^{(l)}_{i,t}{X}^{(m)}_{j,t}\right] \right\rVert <\bar{\varepsilon} \mbox{ for all } l,m \in [p]
	\end{align*}
	implies that
	\begin{equation*}
		\left\lVert \left(\frac{1}{NT}\sum_{i=1}^N\sum_{t=1}^T X_{it}X_{it}'\right)^{-1}\right\rVert 
			\leq 2\left\lVert \left(\frac{1}{NT}\sum_{i=1}^N\sum_{t=1}^T E\left[X_{it}X_{it}'\right]\right)^{-1}\right\rVert 
			\leq \frac{2}{\lambda_\text{min}}~.
	\end{equation*}
	By the union bound plus $\frac{p(p-1)}{2} \times N$ applications of the Bernstein inequality, we have that
	\begin{align*}
		P\left(	\left\lVert \left(\frac{1}{NT}\sum_{i=1}^N\sum_{t=1}^T X_{it}X_{it}'\right)^{-1}\right\rVert > \frac{2}{\lambda_\text{min}} \right)  & \leq \sum_{i = 1}^N  P\left(\left\lVert \left(\frac{1}{T} \sum_{t=1}^T X_{it}X_{it}'\right)^{-1}\right\rVert > \frac{2}{\lambda_\text{min}} \right) \\
		& \leq \, N\frac{p(p-1)}{2} O\left(\sqrt{T} \exp\left( -C_7T \right)  \right) \\
		& \quad + N\frac{p(p-1)}{2} O \left(T\exp\left( -C_8{T}^{\kappa/2} \right)  \right)~.
	\end{align*}	
	where dependence on $\bar{\varepsilon}$ has again been suppressed since we will be treating it as a constant. 
	
	Putting the terms together, 
	\begin{align*}
		& P\left( \left\lVert \hat{\beta} - \beta \right\rVert > \frac{2}{\lambda_\text{min}}p\varepsilon \right) \\
		& \quad \leq N \cdot O\left(\sqrt{T-L} \exp\left( -C_3(T-L)\varepsilon^2 \right)  \right) + N \cdot O \left(\frac{(T-L)}{\varepsilon} \exp\left( -C_4({T-L})^{\kappa/2} \right)  \right) \\ 
		& \quad + N\frac{p(p-1)}{2} O\left(\sqrt{T} \exp\left( -C_7T \right)  \right) + N\frac{p(p-1)}{2} O \left(T\exp\left( -C_8{T}^{\kappa/2} \right)  \right)
	\end{align*}
	Since $\varepsilon$ is arbitrary and $\lambda_\text{min}$ does not change with $N$ or $T$, we are done.

\subsubsection{Proof of Lemma \ref{lemma--closenesstilde}}
We consider the summands individually. 
%\begin{proof}
\begin{align*}
	\left\lvert \tilde{\sigma}_{i,j}^{a,b} - {\sigma}_{i,j}^{a,b} \right\rvert & \leq  \sum_{h=0}^{L} \left( 1- \omega(h, L)\right) \left(	\left\lvert E[X_{i,t}^{(a)}{U}_{i,t}X_{j,t-h}^{(b)}{U}_{j,t-h}] \right\rvert + \left\lvert E[X_{i,t-h}^{(a)}{U}_{i,t-h}X_{j,t}^{(b)}{U}_{j,t}] \right\rvert \right)  \\ 
	& \quad +  \sum_{h=L+1}^{\infty} \left\lvert E[X_{i,t}^{(a)}{U}_{i,t}X_{j,t-h}^{(b)}{U}_{j,t-h}] \right\rvert + \left\lvert E[X_{i,t-h}^{(a)}{U}_{i,t-h}X_{j,t}^{(b)}{U}_{j,t}] \right\rvert ~.
\end{align*}
For the first term, first consider:
\begin{align*}
	 & \sum_{h=0}^{L} \left( 1- \omega(h, L)\right) \left(	\left\lvert E[X_{i,t}^{(a)}{U}_{i,t}X_{j,t-h}^{(b)}{U}_{j,t-h}] \right\rvert \right) \\
	  & \leq \, \sum_{h=0}^{L} \left( 1- \omega(h, L)\right) \left( 12 \cdot \alpha(h)^{1/3} E\left[ \left\lvert X_{i,t}^{(a)}{U}_{i,t} \right\rvert^3 \right]^{1/3} E\left[ \left\lvert X_{j,t-h}^{(b)}{U}_{j,t-h} \right\rvert^3 \right]^{1/3} \right) \\
	  & \leq \, 12 M_3^{1/3}\sum_{h=0}^{L} \left( 1- \omega(h, L)\right) \exp\left(-\frac{C_1}{3}h^\kappa  \right) \\
	  & \to \, 0 \mbox{ as }  L \to \infty
\end{align*}
The first inequality above follows from Davydov's (1968) inequality. The second follows from assumptions \ref{assumption--strongmixing} and \ref{assumption--uniformity}. The final limit follows from the Dominated Convergence Theorem because $\omega(h, L) \to 1^-$ as $L \to \infty$ and $\sum_{h = 0}^\infty \exp\left(-\frac{2C_1}{3}h^\kappa  \right) < \infty$. Hence, for any $\varepsilon$, we can find $L$ large enough so that the first term is smaller than $\varepsilon$. Importantly, this $L$ does not depend on $a,b,i$ or $j$. 

Similarly, for the second term, 
\begin{align*}
	  & \sum_{h=L+1}^{\infty} 	\left\lvert E[X_{i,t}^{(a)}{U}_{i,t}X_{j,t-h}^{(b)}{U}_{j,t-h}] \right\rvert  \\
	  & \leq \, 12 \, \sum_{h= L + 1}^{\infty} \alpha(h)^{1/3} E\left[ \left\lvert X_{i,t}^{(a)}{U}_{i,t} \right\rvert^3 \right]^{1/3} E\left[ \left\lvert X_{j,t-h}^{(b)}{U}_{j,t-h} \right\rvert^3 \right]^{1/3} \\
	  & \leq \, 12 M_3^{1/3}\sum_{h= L + 1}^{\infty} \exp\left(\frac{C_1}{3}h^\kappa  \right) \to \, 0 \mbox{ as }  L \to \infty
\end{align*}
where the first inequality follows from Davydov's inequality and the second follows from assumptions \ref{assumption--strongmixing} and \ref{assumption--uniformity}. We again obtain that for any $\varepsilon$, we can find $L$ independent of $a,b,i,$ and $j$ so that the second term is smaller than $\varepsilon$.

Since our bounds hold uniformly across $a,b,i$ and $j$, we conclude that
\begin{equation*}
	\max_{i,j \in [N]} |\tilde{\sigma}_{i,j} - {\sigma}_{i,j}| \leq \max_{i,j \in [N]} \sum_{a = 1}^d \sum_{b = 1}^d|\tilde{\sigma}^{a,b}_{i,j} - {\sigma}^{a,b}_{i,j}| \to 0 \mbox{ as } L \to \infty~.
\end{equation*}
%\end{proof}

%\end{proof}

\subsubsection{Proof of Lemma \ref{lemma--closenesshat}}

%\begin{proof}
We start by rewriting our term:
\begin{align*}
\left\lvert\hat{\sigma}_{i,j}^{a,b} - \tilde{\sigma}_{i,j}^{a,b} \right\rvert  & \leq \left\lvert \frac{1}{T} \sum_{t=1}^T X_{i,t}^{(a)}\hat{U}_{i,t}X_{j,t}^{(b)}\hat{U}_{j,t} - E[X_{i,t}^{(a)}{U}_{i,t}X_{j,t}^{(b)}{U}_{j,t}] \right\rvert \\
& \quad + \sum_{h=1}^L \omega(h, L)\left\lvert\frac{1}{T-h} \sum_{t = h+1}^T X_{i,t}^{(a)}\hat{U}_{i,t}X_{j,t-h}^{(b)}\hat{U}_{j,t-h} - E[X_{i,t}^{(a)}{U}_{i,t}X_{j,t-h}^{(b)}{U}_{j,t-h}] \right\rvert \\
& \quad + \sum_{h=1}^L \omega(h, L) \left\lvert \frac{1}{T-h} \sum_{t = h+1}^T  X_{i,t}^{(a)}\hat{U}_{i,t}X_{j,t-h}^{(b)}\hat{U}_{j,t-h} - E[X_{i,t}^{(a)}{U}_{i,t}X_{j,t-h}^{(b)}{U}_{j,t-h}]  \right\rvert
\end{align*}
Furthermore,
\begin{align}
\begin{split}
& \left\lvert \frac{1}{T-h} \sum_{t = h+1}^T X_{i,t}^{(a)}\hat{U}_{i,t}X_{j,t-h}^{(b)}\hat{U}_{j,t-h} - E[X_{i,t}^{(a)}{U}_{i,t}X_{j,t-h}^{(b)}{U}_{j,t-h}] \right\rvert \\
& \leq \left\lvert\frac{1}{T-h} \sum_{t=h+1}^T X_{i,t}^{(a)}{U}_{i,t}X_{j,t-h}^{(b)}{U}_{j,t-h} - E[X_{i,t}^{(a)}{U}_{i,t}X_{j,t-h}^{(b)}{U}_{j,t-h}] \right\rvert \\
& \quad + \lVert \hat{\beta} - \beta \rVert \cdot \left\lVert\frac{1}{T-h} \sum_{t=h+1}^T X_{i,t}^{(a)}X_{j,t-h}^{(b)} {X}_{i,t}{U}_{j,t-h} \right\rVert \\
& \quad + \lVert \hat{\beta} - \beta \rVert \cdot \left\lVert\frac{1}{T-h} \sum_{t=h+1}^T X_{i,t}^{(a)}X_{j,t-h}^{(b)} {X}_{i,t-h}{U}_{j,t} \right\rVert \\
& \quad + \lVert \hat{\beta} - \beta \rVert^2 \cdot \left\lVert\frac{1}{T-h} \sum_{t=h+1}^T X_{i,t}^{(a)}X_{j,t-h}^{(b)} {X}_{i,t}{X}_{j,t-h}' \right\rVert
\end{split}\label{eqn--horizondecomp}
\end{align}

Given assumptions \ref{assumption--strongmixing}, \ref{assumption--bernstein} and \ref{assumption--uniformity}, we apply Theorem 1.4 in Bosq (1998) with $q = [\sqrt{T-h}]$ so that:
\begin{multline*}
P\left(\left\lvert\frac{1}{T-h} \sum_{t=h+1}^T X_{i,t}^{(a)}{U}_{i,t}X_{j,t-h}^{(b)}{U}_{j,t-h} - E[X_{i,t}^{(a)}{U}_{i,t}X_{j,t-h}^{(b)}{U}_{j,t-h}] \right\rvert > \varepsilon \right)  \leq \\
\quad a_1 \exp\left(-\frac{\left[\sqrt{T-h}\right]\varepsilon^2}{25M_2^2 + 5C_2 \varepsilon} \right) + a_2 \exp\left( -C_1 \left[\frac{T-h}{\left[\sqrt{T-h}\right]+1}\right]^\kappa  \right)^\frac{2k}{2k+1}
\end{multline*}
where
\begin{align*}
a_1 = 2\frac{T-h}{\left[\sqrt{T-h}\right]} + 2\left( 1 + \frac{\varepsilon^2}{25M_2^2 + 5C_2\varepsilon}\right) \quad \mbox{and} \quad a_2 & = 11 \cdot (T-h) \left(1 + \frac{5 M_k^\frac{k}{2k+1}}{\varepsilon}\right)~.
\end{align*}
Note that for a given $t > s$,  $X_{i,t}^{(a)}{U}_{i,t}X_{j,t-h}^{(b)}{U}_{j,t-h}$ and $ X_{i,s}^{(a)}{U}_{i,s}X_{j,s-h}^{(b)}{U}_{j,s-h}$ are separated by only $(t-s-h)$ periods. Hence, the appropriate bound on the $\alpha$-mixing coefficient is $\exp(-C_1(t-s-h)^{\kappa})$ rather than simply $\exp(-C_1(t-s)^{\kappa})$. However, since $h \leq L \leq \sqrt{T}$, this difference can be simply absorbed into $\kappa$. 

Next, we replace $T-h$ with $T-L$ to obtain a looser upper bound that is uniform across all $h \in [L]$ and $i, j \in [N]$. As $T-L \to \infty$ and $\varepsilon \to 0$, 
\begin{multline}\label{eqn--to0bound}
P\left(\left\lvert\frac{1}{T-h} \sum_{t=h+1}^T X_{i,t}^{(a)}{U}_{i,t}X_{j,t-h}^{(b)}{U}_{j,t-h} - E[X_{i,t}^{(a)}{U}_{i,t}X_{j,t-h}^{(b)}{U}_{j,t-h}] \right\rvert > \varepsilon \right)  \leq \\
\,  O\left(\sqrt{T-L} \exp\left( -C_3(T-L)\varepsilon^2 \right)  \right) + O \left(\frac{(T-L)}{\varepsilon} \exp\left( -C_4({T-L})^{\kappa/2} \right)  \right)
\end{multline}
where
\begin{align*}
C_3 = \frac{1}{25M^2_2} \quad \mbox{and} \quad C_4 & = C_1 \frac{2k}{2k+1}~.
\end{align*}
In fact, the same bound with identical (implicit) constants applies to 
\begin{align*}
\left\lVert\frac{1}{T-h} \sum_{t=h+1}^T X_{i,t}^{(a)}X_{j,t-h}^{(b)}{X}^{(c)}_{i,t}{U}_{j,t-h} \right\rVert \quad \mbox{and} \quad \left\lVert\frac{1}{T-h} \sum_{t=h+1}^T X_{i,t}^{(a)}X_{j,t-h}^{(b)}{X}^{(c)}_{i,t-h}{U}_{j,t} \right\rVert 
\end{align*}
for all $c \in [p]$. Next, since $\lVert \cdot \rVert$ is continuous, there exists $\bar{\varepsilon}$ such that
\begin{align*}
\left\lVert\frac{1}{T-h} \sum_{t=h+1}^T X_{i,t}^{(a)}X_{j,t-h}^{(b)}{X}^{(c)}_{i,t}{X}^{(d)}_{j,t-h} - E\left[X_{i,t}^{(a)}X_{j,t-h}^{(b)}{X}^{(c)}_{i,t}{X}^{(d)}_{j,t-h}\right] \right\rVert <\bar{\varepsilon} \mbox{ for all } c,d \in [p]
\end{align*}
implies that
\begin{equation*}
\left\lVert\frac{1}{T-h} \sum_{t=h+1}^T  X_{i,t}^{(a)}X_{j,t-h}^{(b)}{X}_{i,t}{X}_{j,t-h} \right\rVert \leq 2 \cdot \left\lVert\frac{1}{T-h} \sum_{t=h+1}^T E\left[ X_{i,t}^{(a)}X_{j,t-h}^{(b)}{X}_{i,t}{X}_{j,t-h}\right] \right\rVert \leq 2p^2M_2
\end{equation*}
Applying the argument above to all combinations of $(c,d) \in [p]$, we have that as $T-L \to \infty$, 
\begin{multline*}
P\left(\left\lVert\frac{1}{T-h} \sum_{t=h+1}^T X_{i,t}^{(a)}X_{j,t-h}^{(b)}{X}_{i,t}{X}_{j,t-h} \right\rVert  > 2p^2M_2 \right)  \leq \\
\,  O\left(\sqrt{T-L} \exp\left( -C_5(T-L) \right)  \right) + O \left((T-L) \exp\left( -C_6({T-L})^{\kappa/2} \right)  \right)~.
\end{multline*}	
Note that we have suppressed dependence on $p$ and $M_2$ for convenience. This bound therefore has different implicit constants than in equation  (\ref{eqn--to0bound}). However, the bound is uniform across $h \in [L]$, $a,b \in [p]$ and $i, j \in [N]$. 

Now, fix a $\varepsilon$ and let $A(\varepsilon)$ be the event in which
\begin{itemize}
	\item[i.] $\left\lVert \hat{\beta} - \beta \right\rVert < \varepsilon$, and
	\item[ii.] $
	\left\lvert\frac{1}{T-h} \sum_{t=h+1}^T {W}_{i,t}{W}_{j,t-h} - E[W_{i,t}W_{j,t-h}] \right\rvert < \varepsilon 
	$
	for all $i$, $j$, $h$ , $W_{it} \in \left\{X^{(1)}_{i,t}U_{i,t}, ..., X^{(p)}_{i,t}U_{i,t} \right\}$ and $W_{j,t-h} \in \left\{X^{(1)}_{j,t-h}U_{j,t-h}, ..., X^{(p)}_{j,t-h}U_{j,t-h} \right\}$
	\item[iii.] $\left\lVert\frac{1}{T-h} \sum_{t=h+1}^T X_{i,t}^{(a)}X_{j,t-h}^{(b)}{X}_{i,t}{X}_{j,t-h} \right\rVert  \leq 2p^2M_2$
\end{itemize}
Then on event $A(\varepsilon)$, we have that for all $i,j \in [N]$ and $h \in [L]$, 
\begin{align*}
& \left\lvert \frac{1}{T-h} \sum_{t = h+1}^T X_{i,t}^{(a)}\hat{U}_{i,t}X_{j,t-h}^{(b)}\hat{U}_{j,t-h} - E[X_{i,t}^{(a)}{U}_{i,t}X_{j,t-h}^{(b)}{U}_{j,t-h}] \right\rvert \\
& \quad  \leq \varepsilon + \varepsilon^2 + \varepsilon^2 + 2p^2M_2\varepsilon^2 = \varepsilon + 2(1+p^2M_2)\varepsilon^2~.
\end{align*}
where we substituted all the above bounds into equation (\ref{eqn--horizondecomp}). Taking union bounds over $h \in [L]$, and $a,b \in [p]$ we have that
\begin{align*}
\max_{i,j \in [N]} |\hat{\sigma}_{i,j} - \tilde{\sigma}_{i,j}|  \leq p^2\left(\sum_{h=0}^L \omega(h, L) \right)\varepsilon + 2p^2\left(\sum_{h=0}^L \omega(h, L) \right)(1+p^2M_2)\varepsilon^2
\end{align*}
Choosing $\omega$ to be the Bartlett kernel, by properties of the harmonic series, we have that: 
\begin{equation*}
	L = O(T^{1/2}) \Rightarrow \sum_{h=0}^L \omega(h, L) \leq \log_2\left(T\right)~.
\end{equation*}

By Lemma \ref{lemma--convergenceofbetahat} as well as the calculations above, we know that the event $A^c(\varepsilon)$ -- that is the complement of $A(\varepsilon)$ -- occurs with probability at most:
\begin{align}\label{eqn--sigmahatbound}
& N \cdot O\left(\sqrt{T-L} \exp\left( -C_3(T-L)\varepsilon^2 \right)  \right) + N \cdot O \left(\frac{(T-L)}{\varepsilon} \exp\left( -C_4({T-L})^{\kappa/2} \right)  \right) \\ 
& \quad + N \cdot O\left(\sqrt{T} \exp\left( -C_7T \right)  \right) + N \cdot O \left(T\exp\left( -C_8{T}^{\kappa/2} \right)  \right) \\
& \quad + \frac{N(N+1)}{2} L \cdot O\left(\sqrt{T-L} \exp\left( -C_5(T-L) \right)  \right) + O \left((T-L) \exp\left( -C_6({T-L})^{\kappa/2} \right)  \right) \\
& \quad + \frac{N(N+1)}{2} L \cdot O\left(\sqrt{T-L} \exp\left( -C_3(T-L)\varepsilon^2 \right)  \right) \\
& \quad + \frac{N(N+1)}{2} L \cdot O \left(\frac{(T-L)}{\varepsilon} \exp\left( -C_4({T-L})^{\kappa/2} \right)  \right)
\end{align}
Finally, we let $\varepsilon = T^{-1/2 + \eta}$, $0 < \eta < \frac{1}{2}$. Then with $T - L \to \infty$ and $L \to \infty$, we have that
\begin{equation*}
P\left(A^c(\varepsilon)\right) = N^2 L \cdot O\left(\sqrt{T-L} \exp\left( -C_9(T-L)/L^2 \right)  \right) + N^2 L \cdot O \left({L(T-L)} \exp\left( -C_{10}({T-L})^{\kappa/2} \right)  \right)
\end{equation*}
Given the assumption that $L = o(\sqrt{T})$, we also have that $2p^2L(1+p^2M_2)\varepsilon^2 \to 0$. Furthermore implies that 
\begin{equation*}
P\left(\max_{i,j \in [N]} |\hat{\sigma}_{i,j} - \tilde{\sigma}_{i,j}| > \frac{2 \log_2(T)}{T^{1/2-\eta}}\right) \to 0~.
\end{equation*}
Since $\log_2(T)$ is much smaller than any polynomial rate of $T$, we can eliminate the numerator by increasing $\eta$ slightly. Since $\eta$ is an arbitrary number between $0$ and $1/2$, we are done. 
%\end{proof}

\subsection{Proof of Corollaries \ref{theorem--sizecontrolfixedq} and \ref{theorem--ccesize}}

For any $\nu > 0$, we can find $T$ large enough so that clusters are recovered with probability at least $1-\nu$. 

The assumptions in corollary \ref{theorem--sizecontrolfixedq} ensures that our clusters satisfy the requirements of Theorem 3.1 in \cite{crs2017}. Given the true clusters, approximate randomization test has size $\alpha$. With the estimated clusters, $E[\phi^{CCT}_T] \leq \alpha + \nu$. Since $\nu$ is arbitrary, we are done. 

The assumptions in corollary \ref{theorem--ccesize} ensures that our clusters satisfy the requirements of Theorems 8 and 9 in \cite{hl2019}. As such, with the true clusters, the CCE-based test has size $\alpha$. With the estimated clusters,  $E[\phi^{CCT}_T] \leq \alpha + \nu$. Since $\nu$ is arbitrary, we are done. 

\section{A Simple Illustration of \cite{aaiw17} as a Conditional Treatment Effect}\label{appendix--aaiw}

For ease of exposition, consider a population with $q$ clusters, each with $n$ individuals. We work with the potential outcomes framework where:
\begin{gather*}
	Y_{ij}  = D_{ij}(Y_{ij}(1) - Y_{ij}(0)) + Y_{ij}(0) \\
	Y_{ij}(0) = 0 \quad , \quad Y_{ij}(1) = D_{ij}\beta_j + \varepsilon_{ij} \\
	\beta_j \overset{iid}{\sim} \text{N}(\beta, \sigma_\beta^2) \quad , \quad \varepsilon_{ij} \overset{iid}{\sim} \text{N}(0, \sigma_\varepsilon^2)~.
\end{gather*}
Here, $i \in [n]$ denotes an individual in cluster $j \in [q]$. $D_{ij}$ takes value $1$ if an individual is assigned to treatment and $0$ otherwise. $Y_{ij}(d)$ is the potential outcome of individual $i$ in cluster $j$ under treatment $d$. For any given individual, we only observe one outcome $Y_{ij}$ depending on the treatment assigned. 

Consider the population average treatment effect:
\begin{equation*}
	\text{ATE} = E[Y_{ij}(1) - Y_{ij}(0)] = \beta ~.
\end{equation*}
as well as the ``conditional" average treatment effect:
\begin{equation*}
	\text{CATE} = \frac{1}{q} \sum_{j = 1}^q\beta_j =: \bar{\beta}~.
\end{equation*}
We might think that for policy, what matters is not the true mean but the realized mean for a relevant group. For example, if when studying the effect of minimum wage on unemployment, policy makers are not interested in the mean effect of some abstract data generating process. Instead, they are interested in the mean effect over the 50 states in the US. For this reason, \cite{aaiw17} is places emphasis on CATE. 

\subsection{Clusters Arising From Sample Design}
Consider the following two-step sampling process. First, sample clusters uniformly from $[q]$ without replacement. Denote the set of sampled clusters $\hat{J}$. Then draw all $n$ individuals in each clusters. For each $j$, randomly assign $n/2$ individuals to treatment, and $n/2$ to $0$. Consider the usual OLS estimator:
\begin{equation*}
	\hat{\beta} = \frac{1}{q n/2 } \sum_{j \in \hat{J}}  \sum_{i = 1}^n Y_{ij}D_{ij} -  \frac{1}{q n/2} \sum_{j \in \hat{J}}  \sum_{i = 1}^n Y_{ij}(1-D_{ij})
\end{equation*}
which is just a simple difference-in-means of the treated from untreated individuals. Given our assumptions, we can further write
\begin{align*}
	\hat{\beta} & = \frac{1}{q} \sum_{j \in \hat{J}} \left(\frac{1}{n/2} \sum_{i = 1}^n Y_{ij}D_{ij} -  \frac{1}{n/2} \sum_{i = 1}^n Y_{ij}(1-D_{ij})\right)  \\
	& = \frac{1}{q} \sum_{j \in \hat{J}} \beta_j + \frac{1}{q} \sum_{j \in \hat{J}} \left(\frac{1}{n/2} \sum_{i = 1}^n \varepsilon_{ij}D_{ij} -  \frac{1}{n/2} \sum_{i = 1}^n \varepsilon_{ij}(1-D_{ij})\right)
\end{align*}
so that
\begin{equation*}
	\hat{\beta} - \bar{\beta} \,= \, \underbrace{\frac{1}{q} \sum_{j \in \hat{J}} \beta_j - \bar{\beta}}_\text{clustered}  \,\, + \,\,  \underbrace{\frac{1}{q} \sum_{j \in \hat{J}} \left(\frac{1}{n/2} \sum_{i = 1}^n \varepsilon_{ij}D_{ij} -  \frac{1}{n/2} \sum_{i = 1}^n \varepsilon_{ij}(1-D_{ij})\right)}_{\text{idiosyncratic}}
\end{equation*}
Suppose $\hat{J} = J$. In other words, we have sampled every cluster. Then
\begin{equation*}
	\hat{\beta} - \bar{\beta} = \frac{1}{q} \sum_{j \in {J}} \left(\frac{1}{n/2} \sum_{i = 1}^n \varepsilon_{ij}D_{ij} -  \frac{1}{n/2} \sum_{i = 1}^n \varepsilon_{ij}(1-D_{ij})\right)
\end{equation*}
and the remaining terms vary idiosyncratically with $i$. Hence we are performing inference on the CATE and we observe a representative population, there is no need to cluster the standard errors. This is despite the fact that the regression residuals are correlated. To see this we can rewrite the above model in an estimation equation, we have that:
\begin{equation*}
	Y_{ij} = \underbrace{E[Y_{ij}(0)]}_{\text{$\alpha$}} + D_{ij}\beta + \underbrace{D_{ij}(\beta_j - \beta) + Y_{ij}(0) - E[Y_{ij}(0)]}_{\text{$U_{ij}$}}
\end{equation*}
As such $U_{ij}$ are correlated across individuals because of the heterogeneity $(\beta_j - \beta)$. Nonetheless, this correlation becomes irrelevant once we are considering the asymptotic distribution of $\hat{\beta} - \bar{\beta}$. 

\subsection{Clusters Arising From Experiment Design/Treatment Assignment}
Suppose now that we always sample all individuals in all clusters. In other words, we observe the population. We saw above that if $D_{ij}$ is randomly assigned at the individual level, $\hat{\beta} - \bar{\beta}$ consists only of idiosyncratic terms so that there is no need to cluster. 

Suppose instead that $D_{ij}$ is assigned at the cluster level and that exactly $q/2$ clusters are randomly assigned to treatment $q/2$ are control. Let $J(1)$ and $J(0)$ be the set of treated and control clusters respectively. Then it is easy to see that:
\begin{equation*}
	\hat{\beta} - \bar{\beta} \,= \, \underbrace{\frac{1}{q} \sum_{j \in J(1)} \beta_j - \bar{\beta}}_\text{clustered} \,\, + \,\, 
	\underbrace{\frac{1}{q n/2 } \sum_{j \in J(1)}  \sum_{i = 1}^n \varepsilon_{ij} -  \frac{1}{q n/2} \sum_{j \in {J(0)}}  \sum_{i = 1}^n \varepsilon_{ij}}_\text{idiosyncratic}~,
\end{equation*}
so that again, there would be a need to cluster, this time at the level of treatment assignment. 

We note that in both of the examples above, $\hat{\beta} - \beta$ always involve a term that is clustered. In other words, although researchers interested in CATE only has to cluster if there is the sampling or treatment assignment process leads to an ``unrepresentative sample", researchers interested in ATE always have to cluster the standard errors once the data generating process is clustered.

\end{document}